\documentclass[12pt]{article}

\usepackage{amssymb}
\usepackage{amsmath}
\usepackage{amscd}
\usepackage{graphicx}
\usepackage{subfigure}
\usepackage{rotating}
\usepackage{psfrag}  
\usepackage{xcolor}  
\definecolor{AdinkraGreen}{rgb}{0.10196079, 0.61176473, 0.21960784 }
\definecolor{AdinkraViolet}{rgb}{0.42352942, 0.15294118, 0.4509804 }
\definecolor{AdinkraOrange}{rgb}{0.89803922, 0.57647061, 0.27450982}
\definecolor{AdinkraRed}{rgb}{0.78431374, 0, 0.12156863}
\usepackage{tikz}  
\usetikzlibrary{shapes.misc}  
\usepackage{hyperref}
\usepackage{GatesHeaderIAWilliamsCollege}
\makeatletter
  \newcommand\supertiny{\@setfontsize\supertiny{1pt}{1}}
\makeatother

\begin{document}
\numberwithin{equation}{section}
\setcounter{equation}{0}
\setcounter{page}{0}

\def\dt#1{\on{\hbox{\rm .}}{#1}}                
\def\Dot#1{\dt{#1}}

\def\gfrac#1#2{\frac {\scriptstyle{#1}}
        {\mbox{\raisebox{-.6ex}{$\scriptstyle{#2}$}}}}
\def\gg{{\hbox{\sc g}}}
\border\headpic {\hbox to\hsize{\today \hfill  
{UMDEPP 11-019} 
}}
\par {$~$ \hfill 
{arXiv:1112.2147 [hep-th]} 
} 
\par 

\setlength{\oddsidemargin}{0.3in}
\setlength{\evensidemargin}{-0.3in}

\begin{center}
\vglue .10in
{\large\rm 4D, ${\cal N}$ = 1 Supersymmetry Genomics (II)}\\[.5in]

S.\, James Gates, Jr.\footnote{gatess@wam.umd.edu}${}^{\dagger}$, Jared Hallett
\footnote{jdh4@williams.edu}${}^{\ddagger}$, 
James Parker\footnote{jp@jamesparker.me}${}^{\dagger}$,\\~Vincent G. J. Rodgers\footnote{vrodgers@newton.physics.uiowa.edu}${}^*$,~and Kory Stiffler\footnote{kstiffle@gmail.com}${}^{\dagger}$
\\[0.3in]
${}^\dag${\it Center for String and Particle Theory\\
Department of Physics, University of Maryland\\
College Park, MD 20742-4111 USA}
\\[0.2in]
${}^{\ddagger}${\it Department of Mathematics\\ 
Williams College\\
Williamstown, MA 01267 USA}
\\[0.1in]
{\it {and}}
\\[0.1in]
${}^*${\it Department of Physics and Astronomy\\
The University of Iowa\\
Iowa City, IA 52242 USA}
\\[.4in]
{\rm ABSTRACT}\\[.01in]
\end{center}
\begin{quotation}
{We continue the development of a theory of off-shell supersymmetric representations 
analogous to that of compact Lie algebras such as $SU(3)$.  For off-shell $4D$, ${
\mathcal N}=1$ systems, quark-like representations have been identified~\cite{
Gates:2009me} in terms of cis-Adinkras and trans-Adinkras and it has been 
conjectured that arbitrary  representations are composites of $n_c$-cis  and 
$n_t$-trans representations.  Analyzing the real scalar and complex linear 
superfield multiplets, these ``chemical enantiomer'' numbers are
found to be $n_c = n_t = 1$ and $n_c = 1, n_t=2 $, respectively. }

\endtitle

\setlength{\oddsidemargin}{0.3in}
\setlength{\evensidemargin}{-0.3in}

\setcounter{equation}{0}

\section{Introduction}
This paper is part of continuing efforts to create a comprehensive representation theory 
of off-shell supersymmetry, a project we refer to as supersymmetric `genomics'~\cite{
Gates:2009me}.   From our perspective building the SUSY representation theory is an 
important undertaking.  Representation theory for compact Lie algebras gives us a 
clear mathematical framework with which to separate matter from force carriers in the 
standard model:  matter transforms in the fundamental representation of the gauge 
group, force carriers in the adjoint.  Perhaps an analogous relationship exists among 
yet to be discovered SUSY particles.  If so, an analogous representation theory of 
graded Lie algebras and their associated SUSY systems would prove quite useful.  
In addition, SUSY representation theory would be useful in building gauge/gravity 
correspondences and finding relationships amongst them.  

There is also the matter of the most general four dimensional superstring or heterotic 
string theory.  Although `compactification' of higher dimensional superstring 
or heterotic string theories have long been pursued, there is no `science' to the notion 
of compactification.  One piece of evidence for this is the fact that episodically 'new' 
methods of compactification appear in the physics literature.  The first widely studied 
method along these lines was 'Calabi-Yau compactification.'  But more recently newer 
methods called 'G-2 structures' and 'non-geometrical compactifications' have appeared.
We believe it is a fair question to ask, ``Is there a way to categorize and know all possible 
methods of compactification?''   Whatever the future brings along these lines, at their 
core lies four dimensional $\cal N$ $=$ 1 supersymmetric models and their representation 
theory. In short, any theory which incorporates supersymmetry could benefit from the existence 
of a SUSY representation theory.

Typically, supersymmetric (SUSY) theories are better understood on-shell than they 
are off-shell.  Today, one case of popular interest is in superstring theory: the AdS/CFT 
correspondence.  The low energy effective field theory for Superstring/M-Theory is well 
known to be an on-shell representation of either a 10 or 11 dimensional supergravity~\cite{
GreenSchwarzWitten:1987v2,Polchinski:1998v2}.  In the case of the AdS/CFT correspondence, 
an application manifests itself in the duality between the type IIB supergravity theory and 
another theory which is best understood on-shell: that of $4D$,~${\mathcal N} = 4$ Super 
Yang-Mills theory~\cite{Maldacena:1997re, Aharony:1999ti}.

The issue of on-shell SUSY being better understood than off-shell SUSY has been long 
standing. Since 1981~\cite{Siegel:1981dx} it has remained a question as to whether or 
not an infinite number of auxiliary fields are needed to close the $4D$,~${\mathcal N} = 
4$ Super Yang-Mills off-shell algebra, 
\be
   \left\{ Q_\alpha^I, Q_\beta^J \right\} = 2 \delta^{IJ} P_{\alpha\beta},
\ee
or if indeed this can be accomplished with a finite set.  To resolve this issue it has been 
quite evident for some time that new tools are needed. Recently, an in depth investigation 
of the central charges and internal symmetries of the  $4D$,~${\mathcal N} =4 $ Super 
Yang-Mills algebra was undertaken~\cite{Gates:2011zv}.  In 1995, evidence was shown 
in~\cite{Gates:1995pw} that the minimal size of the off-shell representation with vanishing 
central charges for  $4D$,~${\mathcal N} = 4$ supersymmetry is 128 bosonic and 128 
fermionic degrees of freedom.    This is related to evidence that the representation theories 
of \emph{all} superalgebras in \emph{any} dimension are fully encoded in the representation 
theory of a corresponding one dimensional theory.  It was, of course, satisfying to see 
that these numbers correspond exactly the numbers in $4D$,~${\mathcal N} = 4$ 
superconformal supergravity~\cite{Bergshoeff:1980is}.

Related to this, the graph theoretic tool of Adinkras were introduced in 2004~\cite{
Faux:2004wb}.  Adinkras are graphic representations of the $1D$ `shadows' of 
supersymmetric representations.  It has been conjectured that these are ubiquitous 
in higher dimensions, an example of which is summarized in Fig.~\ref{fig:VCValise}.  
They are similar to Feynman diagrams and Dynkin diagrams in that they convey a 
large amount of mathematical information in a visual form.  Paying homage to the 
phrase, `a picture is worth a thousand words', it will be seen more literal in this paper 
that in the case of Adinkras, a picture can be worth many, many more equations.

Since their inception, Adinkras have been proposed as a useful way to attack 
the problems of supersymmetrical field theory that nonetheless does not actually
rely on field theory.  In a sense the field theory problems have been mapped 
(we believe with complete fidelity) into a realm of graph theory, coding theory,
and other mathematical formalism that permit a rigorous study of all such problems.
In fact, recently there has been proposed a highly formalized mathematical framework~\cite{Zhang:2011kd} that may allow the investigation (and hopefully resolution) of long unsolved problems in this field.

One place where this tool has been used is in the problem of the over-abundance 
of lower dimensional SUSY systems versus higher dimensional systems.  For example, 
not all one dimensional SUSY systems are dimensional reductions of higher dimensional 
systems. It is straightforward to take the higher dimensional theory and reduce it to one 
dimension; going the other way is not so intuitive.  Recently, such a program of dimensionally 
enhancing Adinkras has been undertaken by Faux, Iga, and Landweber~\cite{Faux:2009rz,
Faux:2009wd}. Other Adinkra applications have been in building supersymmetric actions
~\cite{Doran:2006it}, study of error correcting codes~\cite{Doran:2008kh}, and the main 
topic of this paper, SUSY representation theory~\cite{Gates:2009me}.  Also along these
lines~\cite{Gates:2011mu}, the existence of a stringent criterion that bisects one dimensional
representations into two classes (one class only exist for one dimensional representations
and the other can be either one dimensional or two dimensional representations) has
been identified.

In the first part~\cite{Gates:2009me}, it was shown how valise Adinkras (Adinkras which 
are reduced to one fermion row and one boson row as in Fig.~\ref{fig:VCValise}) can be 
used as an organizational tool for off-shell supersymmetric systems.  For instance, two 
distinct, well known $4D$ off-shell supersymmetric systems, the chiral multiplet (Fig.
~\ref{fig:ChiralValise}) and the vector\footnote{This statement works equally well with 
the tensor multiplet replacing the vector multiplet.} multiplet (Fig.~\ref{fig:VectorValise}), 
were shown to have two distinct valise Adinkras: there are no re-definitions solely on 
bosons or alternately solely on fermions which map one of these to the other. 

\begin{figure}[!ht]
\centering
\subfigure[The cis-Adinkra ($n_c = 1, n_t = 0$).  This is the valise Adinkra for the $4D$, 
${\mathcal N} =1$ Chiral multiplet.]{\label{fig:ChiralValise}\includegraphics[width = 
0.45\columnwidth]{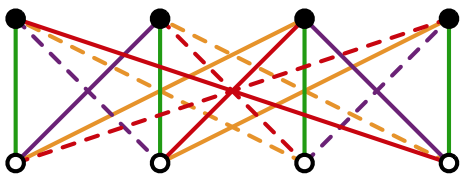}}
\qquad
\subfigure[The trans-Adinkra ($n_c = 0, n_t = 1$).  This is the valise Adinkra for the 
$4D$, ${\mathcal N} =1$ vector and tensor multiplets.]{\label{fig:VectorValise}\includegraphics[width = 0.45\columnwidth]{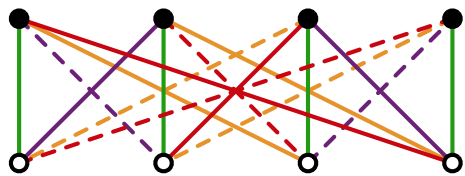}}
\label{fig:VCValise}
\subfigure[One example of chemical enantiomers: the two chiral forms of 
bromo-chloro-fluoro-methane (CHBrClF).]{\psfrag{mirror}[c][c]{mirror}\label{fig:Chiralcarbon}\includegraphics[width=0.8\columnwidth]{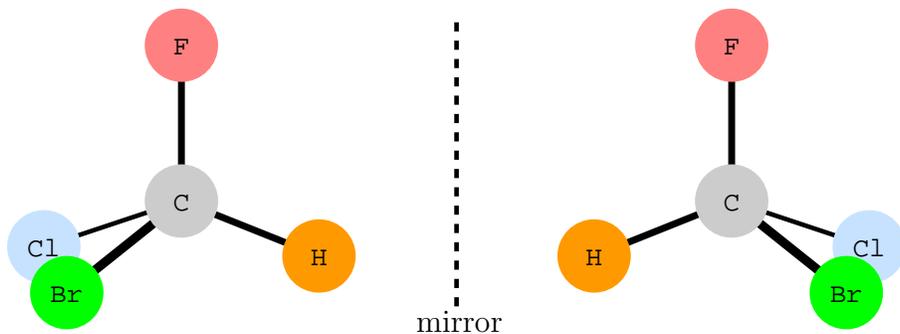}}
\caption{Our conventions for the \subref{fig:ChiralValise} cis- and \subref{fig:VectorValise} 
trans-Adinkra.  They are `color parity reflections' of each other about the `orange axis', 
\subref{fig:Chiralcarbon} analogous to chemical enantiomers which are mirror reflections 
of each other. }
\end{figure}
Notice in Fig.~\ref{fig:VCValise} that a `parity reflection about the orange axis' does 
relate the two diagrams.  More generally, any odd number of such color parity reflections 
will also relate the two.  This is not true of an even number of such reflections.  However, 
it should be noted that color parity reflections correspond to redefinitions of supersymmetry 
generators, not field component redefinitions.  Such `color parity reflections' are analogous 
to spatial reflections of chemical enantiomers, Fig.~\ref{fig:Chiralcarbon}.  Paying homage 
to this analogy, we define the `SUSY enantiomer numbers' $n_c$ and $n_t$ as in Fig.
~\ref{fig:VCValise}.

There is one additional subtlety.  The matter of which of the cis-Adinkra or trans-Adinkra 
is associated with the chiral or vector multiplets is strictly a matter of a choice of conventions.  
As noted in the work of~\cite{Faux:2009rz}, if one begins with both valise Adinkras, it is 
equally valid to assign one to correspond to the chiral multiplet and then necessarily the 
other to the vector multiplet and then to implement a dimensional extension algorithm that 
will be consistent.

It was conjectured that all $4D$, ${\mathcal N} = 1$ off-shell component descriptions of 
supermultiplets are associated with the number of cis-valise ($n_c$) and trans-valise 
($n_t$) Adinkras in the representation.  In this paper, we supply evidence to this conjecture.  
The $4D$, ${\mathcal N} =1$ real scalar superfield off-shell multiplet is found to have the 
SUSY enantiomer numbers $n_c = n_t = 1$, and the $4D$, ${\mathcal N} = 1$ complex 
linear superfield off-shell multiplet is found to have SUSY enantiomer numbers $n_c = 1$ 
and $n_t = 2$.  The discovery of these SUSY enantiomer numbers is also analogous to 
other results much more familiar to particle physicists.  It is well known that for SU(3), the 
dimension ${\bm d}_{SU(3)}$ of {\it {all}} arbitrary representation is specified by two integers 
(we can denote by $p$ and $q$).  For $SU(3)$ the integers $p$ and $q$ also are related
to graphical images called `Young Tableaux' which take the generic form shown in Fig.~\ref{fig:PQReps}.

\begin{figure}[!h]
\centering
\includegraphics[width = 0.6 \columnwidth]{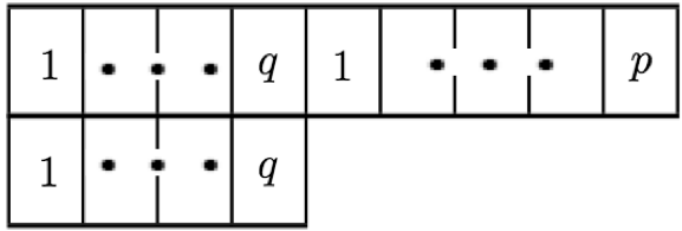}
\caption{The integers $p$ and $q$ in $SU(3)$ Young Tableaux.}
\label{fig:PQReps}
\end{figure}
\noindent where it can be seen that $p$ corresponds to the number of height-1 boxes and
$q$ the number of height-2 boxes in the $SU(3)$ Tableaux. One goal of this SUSY genomics project is to build the representation table for SUSY 
systems in analogy to Lie algebras, as shown in Tab.~\ref{tab:RepTable}.

\begin{table}
  \centering
  \subfigure[$4D$, ${\mathcal N} =1$ SUSY representation table  ]{\label{tab:SUSYRepTable}
  \begin{tabular}{c|c|c|c|c|c}
  \hline
       $n_c$ & 1 & 0 & 1 & 1 & ?\\
       \hline
       $n_t$ & 0 & 1 & 1 & 2 & ? \\
       \hline
       $d$ &  4 & 4 & 8 & 12 & ? \\
       \hline 
       name & C & V or T & RSS & CLS & ?
  \end{tabular}
  }
  \quad
  \subfigure[ $SU(3)$ representation table]{\label{tab:SU3RepTable}
  \begin{tabular}{c|c|c|c|c|c}
                \hline
                $p$ & 1 & 0 & 1 & 3 & \dots \\
                \hline
                $q$ & 0 & 1 & 1 & 0 & \dots \\
                \hline
                $d$   & 3 & 3 & 8 & 10 & \dots \\
                \hline
                name & F & A & O & D & \dots
                \end{tabular}
  }  
  \caption{\subref{tab:SUSYRepTable} The numbers of cis-Adinkras $(n_c)$ and trans-Adinkras $(n_t)$ in the Adinkra representations of the chiral(C), vector(V), tensor(T), real scalar superfield(RSS), and complex linear superfield (CLS) multiplets.  The total number of nodes in each Adinkraic representation is $2d = 8(n_c + n_t)$.  \subref{tab:SU3RepTable} The Lie algebra analogy $SU(3)$ is well known to satisfy the representation size $d = (p+1)(q+1)(p+q+2)/2$.  Some well known particle physics applications are the fundamental(F), anti-fundamental(A), octet(O), and decuplet(D) representations~\cite{Georgi:1999LAPP}. }
  \label{tab:RepTable}
\end{table}

This paper is structured as follows.  Section~\ref{sec:GIReview} reviews parts of genomics part I~\cite{Gates:2009me} pertaining to the current work of part II.  Specifically this includes reviews of the $4D$, ${\mathcal N} = 1$ off-shell chiral and vector multiplets.  New in this work is the specific form of the Lagrangians of these multiplets as well as their zero-brane reduction.  The Adinkras for these multiplets can be found in the supporting extended review in App.~\ref{app:AdinkraReview}.  New in these Adinkras are the explicit nodal field content.

The bulk of the new results of this paper are found in section~\ref{sec:RSM} which introduces the real scalar superfield multiplet and section~\ref{sec:CLM} which introduces the complex linear superfield multiplet.  Both these sections follow the same structure.  First are shown the SUSY transformations laws and the Lagrangians of which they are an invariance.  Following this, their zero-brane reductions are explicitly shown, and their Adinkras are drawn with explicit nodal field content.  In both cases, it is shown how the SUSY enantiomer numbers can be read directly from these Adinkras.  Also, traces referred to in part I~\cite{Gates:2009me} as chromocharacters are derived for each of these multiplets.  Spring boarding off the trace results of Part I, it is shown how the SUSY enantiomer numbers of these two newly investigated multiplets are encoded within these traces.  Additionally in section~\ref{sec:RSM} the real pseudo-scalar supermultiplet and its relation to the real scalar supermultiplet are briefly discussed.

Finally, in section~\ref{sec:CharPs}, we define the characteristic polynomial for Adinkras.  We show how this polynomial encodes the precise numerical discrepancy between bosonic and fermionic nodes in the Adinkra.  All conventions, including those for the gamma matrices, are as in Part I~\cite{Gates:2009me}, unless otherwise specified.

\section{Review of Genomics(I)}\label{sec:GIReview}
In this section, we review the results from Genomics(I)~\cite{Gates:2009me} and pertinent to our current work herein contained.  In~\cite{Gates:2009me}, we studied the well known $4D$, ${\mathcal N} =1 $ chiral multiplet and the $4D$, ${\mathcal N} =1 $ vector multiplet, both in a Majorana spinor representation.  The supersymmetric transformations, algebra, and reductions to the zero-brane were investigated for these representations.  The zero-brane reduction of the supersymmetric transformations led us to the Adinkra representations of these two systems.  In the following two sections, we review the Lagrangians for each system, the zero-brane reduction of this Lagrangian, and the resulting Adinkras which encode the supersymmetric transformation laws for the zero-brane reduced Lagrangians.

\subsection{The \texorpdfstring{$4D$}{4D} \texorpdfstring{${\mathcal N}$ =1}{N=1} Chiral Multiplet}
 The 4D, $\cal N$ = 1 chiral multiplet is very well known to consist of a scalar $A$,
a pseudoscalar $B$, a Majorana fermion $\psi_a$, a scalar auxiliary field $F$, and a
pseudoscalar auxiliary field $G$.  The Lagrangian for this system which is supersymmetric with respect to the transformation laws investigated in~\cite{Gates:2009me} is:
\be\label{eq:CMLagrangian}\eqalign{
   {\mathcal L}_{CM} = & -\frac{1}{2}(\partial_{\mu}A)(\partial^{\mu}A) -\frac{1}{2}(\partial_{\mu} B)(\partial^{\mu}B)\cr
&+i\frac{1}{2}(\gamma^{\mu})^{ab}\psi_{a}\partial_{\mu}\psi_{b} +\frac{1}{2} F^{2}+\frac{1}{2} G^{2} 
}\ee

It's zero-brane reduction is acquired by assuming only time dependence of the fields.  This Lagrangian with time derivatives denoted by a prime $(')$ is:
\be\label{eq:CMLagrangian0}\eqalign{
   {\mathcal L}_{CM}^{(0)}= & \frac{1}{2}(A')^2 +\frac{1}{2}(B')^2 +i\frac{1}{2}\delta^{ab}\psi_{a}\psi_{b}' +\frac{1}{2} F^{2}+\frac{1}{2} G^{2} 
}\ee
The zero-brane reduced SUSY transformation laws that are a symmetry of this Lagrangian were shown in~\cite{Gates:2009me} to be succinctly written as the valise cis-Adinkra in Fig.~\ref{fig:ChiralValise}.  This is reviewed in App.~\ref{app:AdinkraReview}.

\subsection{The \texorpdfstring{$4D$}{4D} \texorpdfstring{${\mathcal N}$ =1}{N=1} Vector Multiplet}
The 4D, $\cal N$ = 1 vector multiplet off-shell is described by a vector
 $A{}_{\mu}$, a Majorana fermion $\l_a$, and a pseudoscalar auxiliary field
 d.  The Lagrangian for this system which is supersymmetric with respect to the transformation laws investigated in~\cite{Gates:2009me} is:
\be\label{eq:VMLagrangian}\eqalign{
   {\mathcal L}_{VM} = &-\frac{1}{4}F_{\mu\nu}F^{\mu\nu} +\frac{1}{2}i(\gamma^{\mu})^{ab}\lambda_{a}\partial_{\mu}\lambda_{b}+\frac{1}{2}{\rm d}^2.
}\ee 
It's zero-brane reduction is acquired by assuming only time dependence of the fields.  This Lagrangian with time derivatives denoted by a prime $(')$ is:
\be\label{eq:VMLagrangian0}\eqalign{
   {\mathcal L}_{VM}^{(0)} = &\frac{1}{2}((A_1')^2 + (A_2')^2 + (A_3')^2) +\frac{1}{2}i \delta^{ab}\lambda_{a}\lambda_{b}'+\frac{1}{2}{\rm d}^2
}\ee 
The zero-brane reduced SUSY transformation laws which are a symmetry of this Lagrangian were shown in~\cite{Gates:2009me} to be succinctly written as the valise trans-Adinkra in Fig.~\ref{fig:VectorValise}.  This is reviewed in App.~\ref{app:AdinkraReview}.

\section{The \texorpdfstring{$4D$}{4D}, \texorpdfstring{$\cal N$ = 1}{N=1} Real Scalar Superfield Multiplet }\label{sec:RSM}

$~~~$ The 4D, $\cal N$ = 1 real scalar superfield is a multiplet that is very well known.  It consists of a scalar $K$,
a Majorana fermion $\zeta$, a scalar field $M$, a pseudoscalar field $N$, and axial vector field  $U$, a
Majorana fermion field $\Lambda$, and another scalar field ${\rm d}$.  All together, their are eight bosonic and eight fermionic degrees.
\subsection{Supersymmetry Transformation Laws}
 We use the following set of transformation laws of the supercovariant derivative, ${\rm D}_a$, acting on each component of the real scalar superfield multiplet:
\be\eqalign{
 { \rm D}_a K= & \z_a \cr
 { \rm D}_a M = & \frac{1}{2}\L_a -\frac{1}{2} (\g^\nu)_{a}^{\,\,\,d} \partial_\nu \z_d \cr
 { \rm D}_a N = &-i \frac{1}{2} (\g^5)_a^{\,\,\,\,d} \L_d +i \frac{1}{2} (\g^5 \g^\nu)_{a}^{\,\,\,d} \partial_\nu \z_d \cr
 { \rm D}_a U_\mu = & i \frac{1}{2}(\g^5 \g_\mu)_a^{\,\,\,\,d} \L_d -i \frac{1}{2} (\g^5 \g^\nu \g_\mu)_{a}^{\,\,\,d} \partial_\nu \z_d \cr
 { \rm D}_a {\rm d} = & -(\g^\nu)_a^{\,\,\,d} \partial_\nu \L_d \cr
 { \rm D}_a \z_b = & i (\g^\mu)_{ab} \partial_\mu K + ( \g^5 \g^\mu)_{ab} U_\mu +i C_{ab} M + (\g^5)_{ab} N \cr
 { \rm D}_a \L_b = & i (\g^\mu)_{ab} \partial_\mu M + (\g^5 \g^\mu)_{ab} \partial_\mu N+ ( \g^5 \g^\mu \g^\nu)_{ab} \partial_\mu U_\nu + i C_{ab} {\rm d}. 
}\label{eq:DRSM}\ee

A direct calculation shows that
\begin{equation}
  \{ { \rm D}_a, { \rm D}_b \} = 2 i (\gamma^\mu)_{ab} \partial_\mu
\end{equation}
is satisfied for all fields in the multiplet   Furthermore, the following Lagrangian is invariant, up to total derivatives, with respect to the $D_a$ transformations in Eq.~(\ref{eq:DRSM}):
\begin{equation}
   {\mathcal L} = -\frac{1}{2} M^2 - \frac{1}{2}N^2 + \frac{1}{2} U_\mu U^\mu - \frac{1}{2} K {\rm d} + i \frac{1}{2}\zeta_a C^{ab} \Lambda_b
\end{equation}
It is clear this Lagrangian implies that all fields in this multiplet are non-propagating.

The real scalar supermultiplet can be used to derive the structure of the
real pseudo-scalar supermultiplet by making the substitutions
$$
\label{substitution1}\begin{array}{cccc}K\rightarrow L & \zeta_a\rightarrow 
i (\gamma^5)_{a}^{\,\,\,d}\rho_d & M\rightarrow\hat{N} & N\rightarrow 
-\hat{M}\end{array}
$$
\begin{equation}\label{substitution2}\begin{array}{ccc}U_\mu\rightarrow V_\mu & \Lambda_a\rightarrow -i(\gamma^5)_{a}^{\,\,\,d}\hat{\Lambda}_d & {\rm d}\rightarrow\hat{{\rm d}} \end{array}\end{equation}
and this yields the pseudo-scalar supermultiplet, which satisfies
\be\eqalign{
{\rm D}_aL= &i (\gamma^5)_{a}^{\,\,\,d}\rho_d \cr
{\rm D}_a\rho_b= &-(\gamma^5\gamma^\mu)_{ab}\partial_\mu L+i(\gamma^\mu)_{ab}V_\mu+(\gamma^5)_{ab}\hat{N}+iC_{ab}\hat{M} \cr
{\rm D}_a\hat{N}=&-i \frac{1}{2} (\gamma^5)_{a}^{\,\,\,d}\hat{\Lambda}_d+i\frac{1}{2} (\gamma^5\gamma^\mu)_{a}^{\,\,\,d}\partial_\mu\rho_d \cr
{\rm D}_a\hat{M}= &\frac{1}{2} \hat{\Lambda}_a-\frac{1}{2} (\gamma^\mu)_{a}^{\,\,\,d}\partial_\mu\rho_d \cr
{\rm D}_aV_\mu= &- \frac{1}{2} (\gamma_\mu)_{a}^{\,\,\,d}\hat{\Lambda}_d+\frac{1}{2} (\gamma^\nu\gamma_\mu)_{a}^{\,\,\,d}\partial_\nu\rho_d \cr
{\rm D}_a\hat{\rm d}= & -i (\gamma^5\gamma^\mu)_{a}^{\,\,\,d}\partial_\mu\hat{\Lambda}_d \cr
{\rm D}_a\hat{\Lambda}_b=&(\gamma^5\gamma^\mu)_{ab}\partial_\mu\hat{N}+i(\gamma^\mu)_{ab}\partial_\mu\hat{M}+i(\gamma^\mu\gamma^\nu)_{ab}\partial_\mu V_\nu-(\gamma^5)_{ab}\hat{\rm d}
}\ee

\subsection{One Dimensional Reduction}
On our way to the Adinkra picture of the real scalar superfield multiplet, we here first reduce the transformation laws, Eq.~(\ref{eq:DCLM1}), to one dimension by considering the fields to have only time dependence.  In the following, we list these time-only-dependent transformation laws, with time derivatives denoted by a prime $(')$.  The transformation laws on the bosons are

\begin{equation}
\begin{array}{llll}
{ \rm D}_1 K= \z_1  & { \rm D}_2 K=\z_2  &
{ \rm D}_3 K=\z_3 & { \rm D}_4 K = \z_4   
\end{array}
\end{equation}
\begin{equation}
\begin{array}{ll}
{ \rm D}_1 M= \frac{1}{2}\L_1-\frac{1}{2}\z'_2 & { \rm D}_2 M=\frac{1}{2}\L_2+\frac{1}{2}\z'_1  \\
 { \rm D}_3 M=\frac{1}{2}\L_3+\frac{1}{2}\z'_4  & { \rm D}_4 M = \frac{1}{2}\L_4-\frac{1}{2}\z'_3   \end{array}
 \end{equation}
\begin{equation}
\begin{array}{ll}
{ \rm D}_1 N= \frac{1}{2}\L_4-\frac{1}{2}\z'_3   & { \rm D}_2 N=-\frac{1}{2}\L_3-\frac{1}{2}\z'_4  \\
 { \rm D}_3 N=\frac{1}{2}\L_2+\frac{1}{2}\z'_1   & { \rm D}_4 N = -\frac{1}{2}\L_1+\frac{1}{2}\z'_2  \end{array}
 \end{equation}
\begin{equation}
\begin{array}{llll}
 { \rm D}_1 {\rm d} = -\L'_2  & { \rm D}_2 {\rm d} = \L'_1  & { \rm D}_3 {\rm d} = \L'_4  & { \rm D}_4 {\rm d} = -\L'_3 
 \end{array}
 \end{equation}
\begin{equation}
 \begin{array}{ll}
{ \rm D}_1 U_0 = \frac{1}{2}\L_3+\frac{1}{2} \z'_4 & { \rm D}_1 U_1 = \frac{1}{2} \zeta'_4-\frac{1}{2}\Lambda_3 \\ { \rm D}_1 U_2 =  \frac{1}{2}\Lambda_1
 +\frac{1}{2} \zeta'_2 &{ \rm D}_1 U_3 =  \frac{1}{2}\Lambda_4+\frac{1}{2} \zeta'_3 
 \end{array}
 \end{equation}
 \begin{equation}
 \begin{array}{ll}
{ \rm D}_2 U_0 =  \frac{1}{2}\Lambda_4-\frac{1}{2} \zeta'_3 &{ \rm D}_2 U_1 =  \frac{1}{2}\Lambda_4+\frac{1}{2} \zeta'_3 \\
 { \rm D}_2 U_2 =  \frac{1}{2}\Lambda_2-\frac{1}{2} \z'_1 &{ \rm D}_2 U_3 =  \frac{1}{2}\Lambda_3-\frac{1}{2} \zeta'_4
\end{array}
 \end{equation}
 \begin{equation}
 \begin{array}{ll} 
{ \rm D}_3 U_0 =  \frac{1}{2} \zeta'_2-\frac{1}{2}\Lambda_1 
&{ \rm D}_3 U_1 =  -\frac{1}{2}\Lambda_1-\frac{1}{2} \zeta'_2 \\
 { \rm D}_3 U_2 =  \frac{1}{2} \zeta'_4-\frac{1}{2}\Lambda_3 &
 { \rm D}_3 U_3 =  \frac{1}{2}\Lambda_2-\frac{1}{2} \zeta'_1 
 \end{array}
 \end{equation}
 \begin{equation}
 \begin{array}{ll}  
{ \rm D}_4 U_0 =  -\frac{1}{2}\Lambda_2-\frac{1}{2} \zeta'_1 &
 { \rm D}_4 U_1 =  \frac{1}{2}\Lambda_2-\frac{1}{2} \zeta'_1 \\
 { \rm D}_4 U_2 =  -\frac{1}{2}\Lambda_4-\frac{1}{2} \zeta'_3 & { \rm D}_4 U_3 =  \frac{1}{2}\Lambda_1+\frac{1}{2} \zeta'_2
 \end{array}
\end{equation}
 and on the fermions are
\begin{equation}
 \begin{array}{ll}
{ \rm D}_1 \z_1=  i K' &{ \rm D}_1 \z_2=  i U_2-i M \\ { \rm D}_1 \z_3=  i U_3-i N & { \rm D}_1 \z_4=  i U_0+i U_1  
\end{array}
\end{equation}
\begin{equation}
\begin{array}{ll}
{ \rm D}_2 \z_1 = i M-i U_2 &{ \rm D}_2 \z_2=  i K' \\
 { \rm D}_2 \z_3 =  i U_1-i U_0 &{ \rm D}_2 \z_4=  -i N-i U_3 
 \end{array}
\end{equation}
\begin{equation}
\begin{array}{ll}
{ \rm D}_3 \z_1=  i N-i U_3 &{ \rm D}_3 \z_2=  i U_0-i U_1 \\
 { \rm D}_3 \z_3=  i K' &{ \rm D}_3 \z_4=  i M+i U_2 
 \end{array}
\end{equation}
\begin{equation}
\begin{array}{ll}
{ \rm D}_4 \z_1=  -i U_0-i U_1 & { \rm D}_4 \z_2=  i N+i U_3 \\ { \rm D}_4 \z_3=  -i M-i U_2 &{ \rm D}_4 \z_4=  i K'
\end{array}
\end{equation}
\begin{equation}
\begin{array}{llll}
{ \rm D}_1 \L_1=   i U'_2+ i M' &{ \rm D}_1 \L_2=   -i d  \\ { \rm D}_1 \L_3=  i U'_0- i U'_1 & { \rm D}_1 \L_4=   i N'+ i U'_3 
\end{array}
\end{equation}
\begin{equation}
\begin{array}{ll}
   { \rm D}_2 \L_1=   i {\rm d} &{ \rm D}_2 \L_2=  i U'_2+ i M' \\ 
   { \rm D}_2 \L_3=  i U'_3-  i N' &{ \rm D}_2 \L_4=   i U'_0+ i U'_1 
   \end{array}
\end{equation}
\begin{equation}
\begin{array}{ll}
{ \rm D}_3 \L_1=  -i U'_0- i U'_1 &{ \rm D}_3 \L_2=   i N'+ i U'_3 \\
 { \rm D}_3 \L_3=   i M'- i
   U'_2 & { \rm D}_3 \L_4=   i {\rm d} 
   \end{array}
\end{equation}
\begin{equation}
\begin{array}{ll}
{ \rm D}_4 \L_1=   i U'_3- i N' &{ \rm D}_4 \L_2 =   i U'_1-  i U'_0 \\
{ \rm D}_4 \L_3=  -i {\rm d} &{ \rm D}_4 \L_4=  i M'-  i U'_2~~~.
\end{array}
\end{equation}
These transformation laws can be depicted as the Adinkra in Fig.~\ref{fig:RSFAdinkra1}.  The rules for drawing Adinkras are reviewed in app.~\ref{app:CM}.

\begin{figure}[!h]
  \centering
    \psfrag{b8}[c][c]{\fcolorbox{black}{white}{$K$}}
    \psfrag{f7}[c][c]{\begin{tikzpicture}
 \node[rounded rectangle,draw,fill=white!30]{$\zeta_1$};
 \end{tikzpicture}}
    \psfrag{f6}[c][c]{\begin{tikzpicture}
 \node[rounded rectangle,draw,fill=white!30]{$\zeta_2$};
 \end{tikzpicture}}
    \psfrag{f5}[c][c]{\begin{tikzpicture}
 \node[rounded rectangle,draw,fill=white!30]{$\zeta_3$};
 \end{tikzpicture}}
    \psfrag{f8}[c][c]{\begin{tikzpicture}
 \node[rounded rectangle,draw,fill=white!30]{$\zeta_4$};
 \end{tikzpicture}}
    \psfrag{b2}[c][c]{\fcolorbox{black}{white}{$U_0 - U_1$}}
    \psfrag{b3}[c][c]{\fcolorbox{black}{white}{$U_3 - N$}}
    \psfrag{b4}[c][c]{\fcolorbox{black}{white}{$M + U_2$}}
    \psfrag{b5}[c][c]{\fcolorbox{black}{white}{ $M - U_2$}}
    \psfrag{b6}[c][c]{\fcolorbox{black}{white}{$U_3 + N$}}
    \psfrag{b7}[c][c]{\fcolorbox{black}{white}{$U_0 + U_1$}}
    \psfrag{f3}[c][c]{\begin{tikzpicture}
 \node[rounded rectangle,draw,fill=white!30]{$\Lambda_1$};
 \end{tikzpicture}}
    \psfrag{f2}[c][c]{\begin{tikzpicture}
 \node[rounded rectangle,draw,fill=white!30]{$\Lambda_2$};
 \end{tikzpicture}}
    \psfrag{f1}[c][c]{\begin{tikzpicture}
 \node[rounded rectangle,draw,fill=white!30]{$\Lambda_3$};
 \end{tikzpicture}}
    \psfrag{f4}[c][c]{\begin{tikzpicture}
 \node[rounded rectangle,draw,fill=white!30]{$\Lambda_4$};
 \end{tikzpicture}} 
    \psfrag{b1}[c][c]{\fcolorbox{black}{white}{${\rm d}$}}
  \includegraphics[width =  \columnwidth]{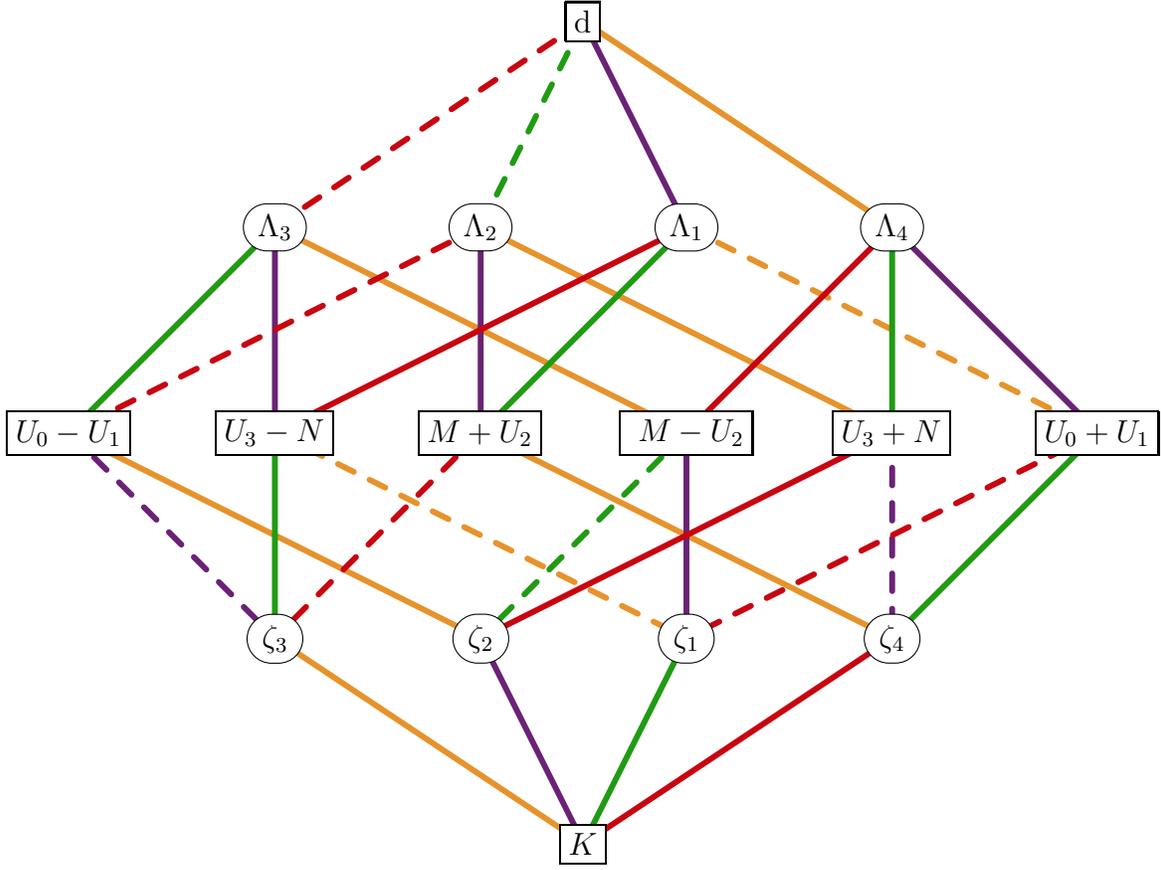}
 \caption{An Adinkra for the real scalar superfield multiplet.  Color convention: \textcolor{AdinkraGreen}{${ \rm D}_1$}, \textcolor{AdinkraViolet}{${ \rm D}_2$}, \textcolor{AdinkraOrange}{${ \rm D}_3$}, \textcolor{AdinkraRed}{${ \rm D}_4$}. }
\label{fig:RSFAdinkra1}
\end{figure}

\subsection{Canonical Fields, \texorpdfstring{$\Phi$}{P} and \texorpdfstring{$\Psi$}{P}}
Defining the nodes canonically from the Adinkra, fig.~\ref{fig:RSFAdinkra1}, as
$$
   \Phi_i = \left(
     \begin{array}{c}
      \int {\rm d}~dt  \\
      M + U_2 \\
      U_0 + U_1 \\
      N+U_3 \\
     -N + U_3 \\
      U_0 - U_1 \\
     -M + U_2 \\
      K' 
     \end{array}
     \right),~~~    
  i \Psi_{\hat{i}} =  \left( 
     \begin{array}{c}
      \L_1 \\
      \L_2 \\
      \L_3 \\
      \L_4 \\
      \z_1 ' \\
      \z_2 ' \\
      \z_3 ' \\
      \z_4 '
     \end{array}
    \right)
$$
we can write the transformation laws succinctly as
\be\label{eq:AdinkraMatrixDefinition}
   { \rm D}_{\rm I} \Phi_i = i ( {\rm L }_{\rm I})_{i\hat{j}} \Psi_{\hat{j}},~~~{ \rm D}_{\rm I} \Psi_{\hat{j}} = ({\rm R }_{\rm I})_{\hat{j}i}\Phi_i'
\ee
where the Adinkra matrices
\be
{ \bm {\rm L}}_1 = \left(
\begin{array}{cccccccc}
 0 & -1 & 0 & 0 & 0 & 0 & 0 & 0 \\
 1 & 0 & 0 & 0 & 0 & 0 & 0 & 0 \\
 0 & 0 & 0 & 0 & 0 & 0 & 0 & 1 \\
 0 & 0 & 0 & 1 & 0 & 0 & 0 & 0 \\
 0 & 0 & 0 & 0 & 0 & 0 & 1 & 0 \\
 0 & 0 & 1 & 0 & 0 & 0 & 0 & 0 \\
 0 & 0 & 0 & 0 & 0 & 1 & 0 & 0 \\
 0 & 0 & 0 & 0 & 1 & 0 & 0 & 0
\end{array}
\right),
\ee

\be
{ \bm {\rm L}}_2 = \left(
\begin{array}{cccccccc}
 1 & 0 & 0 & 0 & 0 & 0 & 0 & 0 \\
 0 & 1 & 0 & 0 & 0 & 0 & 0 & 0 \\
 0 & 0 & 0 & 1 & 0 & 0 & 0 & 0 \\
 0 & 0 & 0 & 0 & 0 & 0 & 0 & -1 \\
 0 & 0 & 1 & 0 & 0 & 0 & 0 & 0 \\
 0 & 0 & 0 & 0 & 0 & 0 & -1 & 0 \\
 0 & 0 & 0 & 0 & -1 & 0 & 0 & 0 \\
 0 & 0 & 0 & 0 & 0 & 1 & 0 & 0
\end{array}
\right),
\ee

\be
{ \bm {\rm L}}_3 = \left(
\begin{array}{cccccccc}
 0 & 0 & 0 & 1 & 0 & 0 & 0 & 0 \\
 0 & 0 & 0 & 0 & 0 & 0 & 0 & 1 \\
 -1 & 0 & 0 & 0 & 0 & 0 & 0 & 0 \\
 0 & 1 & 0 & 0 & 0 & 0 & 0 & 0 \\
 0 & 0 & 0 & 0 & -1 & 0 & 0 & 0 \\
 0 & 0 & 0 & 0 & 0 & 1 & 0 & 0 \\
 0 & 0 & -1 & 0 & 0 & 0 & 0 & 0 \\
 0 & 0 & 0 & 0 & 0 & 0 & 1 & 0
\end{array}
\right),
\ee

\be
{ \bm {\rm L}}_4 = \left(
\begin{array}{cccccccc}
 0 & 0 & -1 & 0 & 0 & 0 & 0 & 0 \\
 0 & 0 & 0 & 0 & 0 & 0 & -1 & 0 \\
 0 & 0 & 0 & 0 & -1 & 0 & 0 & 0 \\
 0 & 0 & 0 & 0 & 0 & 1 & 0 & 0 \\
 1 & 0 & 0 & 0 & 0 & 0 & 0 & 0 \\
 0 & -1 & 0 & 0 & 0 & 0 & 0 & 0 \\
 0 & 0 & 0 & -1 & 0 & 0 & 0 & 0 \\
 0 & 0 & 0 & 0 & 0 & 0 & 0 & 1
\end{array}
\right),
\ee
satisfy the orthogonal relationship

\be
  {\bm {\rm R}}_{\rm I} = { \bm {\rm L}}_{\rm I}^t = { \bm {\rm L}}_{\rm I}^{-1} ,
\ee
and satisfy the garden algebra
\be\eqalign{
   { \bm {\rm L}}_{\rm I} {\bm {\rm R}}_{\rm J} + { \bm {\rm L}}_{\rm J} {\bm {\rm R}}_{\rm I} = 2 \delta_{\rm IJ} {\bm {\rm I}}_8 \cr
   {\bm {\rm R}}_{\rm I} { \bm {\rm L}}_{\rm J} + {\bm {\rm R}}_{\rm J} { \bm {\rm L}}_{\rm I} = 2 \delta_{\rm IJ} {\bm {\rm I}}_8
}\ee
with ${\bm {\rm I}}_n$ the $n \times n$ identity matrix.

\subsection{Finding \texorpdfstring{$n_c$}{n} and \texorpdfstring{$n_t$}{n} via Adinkras.}\label{sec:RSSO8Transforms}
Applying the field redefinitions
\be
   \hat{\Phi}' = {\bm {\cal X}} \Phi = 
   \frac{1}{\sqrt{2}}
   \left(   
   \begin{array}{c}
       \int {\rm d}~dt - K' \\
       2 M \\
       2 N \\
       2 U_0 \\
       -2 U_2 \\
       K' + \int {\rm d}~dt\\
       2 U_1 \\
       -2 U_3                           
   \end{array}
   \right),~~~
   \hat{\Psi}' = {\bm {\mathcal Y}}^t\Psi = 
   i\frac{1}{\sqrt{2}}
   \left( 
   \begin{array}{c}
        \z_1' + \L_2 \\
        \z_2' - \L_1 \\
        \z_3' - \L_4 \\
        -\z_4' - \L_3 \\
        \z_2' + \L_1 \\
        -\z_1' + \L_2 \\
        -\z_4' + \L_3 \\
        \z_3' + \L_4
   \end{array}
   \right)                  
\ee
where the matrices are elements of the $O(8)$ group and given by
\be\label{eq:calX8}
\bm{ {\cal X}} = \frac{1}{\sqrt{2}}\left(
\begin{array}{cccccccc}
 1 & 0 & 0 & 0 & 0 & 0 & 0 & -1 \\
 0 & 1 & 0 & 0 & 0 & 0 & -1 & 0 \\
 0 & 0 & 0 & 1 & -1 & 0 & 0 & 0 \\
 0 & 0 & 1 & 0 & 0 & 1 & 0 & 0 \\
 0 & -1 & 0 & 0 & 0 & 0 & -1 & 0 \\
 1 & 0 & 0 & 0 & 0 & 0 & 0 & 1 \\
 0 & 0 & 1 & 0 & 0 & -1 & 0 & 0 \\
 0 & 0 & 0 & -1 & -1 & 0 & 0 & 0
\end{array}
\right),
\ee

\be\label{eq:calY8}
{\bm {\mathcal Y}} = \frac{1}{\sqrt{2}}\left(
\begin{array}{cccccccc}
 0 & 1 & 0 & 0 & -1 & 0 & 0 & 0 \\
 -1 & 0 & 0 & 0 & 0 & -1 & 0 & 0 \\
 0 & 0 & 0 & 1 & 0 & 0 & -1 & 0 \\
 0 & 0 & 1 & 0 & 0 & 0 & 0 & -1 \\
 -1 & 0 & 0 & 0 & 0 & 1 & 0 & 0 \\
 0 & -1 & 0 & 0 & -1 & 0 & 0 & 0 \\
 0 & 0 & -1 & 0 & 0 & 0 & 0 & -1 \\
 0 & 0 & 0 & 1 & 0 & 0 & 1 & 0
\end{array}
\right),
\ee
take us to another Adinkraic basis via
\be
   \hat{{ \bm {\rm L}}}_{\rm I} = {\bm {\cal X}} { \bm {\rm L}}_{\rm I} {\bm {\mathcal Y}},~~~\hat{{\bm {\rm R}}}_{\rm I} = {\bm {\mathcal Y}}^t {\bm {\rm R}}_{\rm I} {\bm {\cal X}}^t
\ee
where the Adinkra matrices now are block diagonal:
\be
\hat{{ \bm {\rm L}}}_1 = \left(
\begin{array}{cccccccc}
 1 & 0 & 0 & 0 & 0 & 0 & 0 & 0 \\
 0 & 1 & 0 & 0 & 0 & 0 & 0 & 0 \\
 0 & 0 & 1 & 0 & 0 & 0 & 0 & 0 \\
 0 & 0 & 0 & 1 & 0 & 0 & 0 & 0 \\
 0 & 0 & 0 & 0 & 1 & 0 & 0 & 0 \\
 0 & 0 & 0 & 0 & 0 & 1 & 0 & 0 \\
 0 & 0 & 0 & 0 & 0 & 0 & 1 & 0 \\
 0 & 0 & 0 & 0 & 0 & 0 & 0 & 1
\end{array}
\right),
\ee

\be
  \hat{{ \bm {\rm L}}}_2 = \left(
\begin{array}{cccccccc}
 0 & 1 & 0 & 0 & 0 & 0 & 0 & 0 \\
 -1 & 0 & 0 & 0 & 0 & 0 & 0 & 0 \\
 0 & 0 & 0 & -1 & 0 & 0 & 0 & 0 \\
 0 & 0 & 1 & 0 & 0 & 0 & 0 & 0 \\
 0 & 0 & 0 & 0 & 0 & 1 & 0 & 0 \\
 0 & 0 & 0 & 0 & -1 & 0 & 0 & 0 \\
 0 & 0 & 0 & 0 & 0 & 0 & 0 & -1 \\
 0 & 0 & 0 & 0 & 0 & 0 & 1 & 0
\end{array}
\right),
\ee

\be
  \hat{{ \bm {\rm L}}}_3 = \left(
\begin{array}{cccccccc}
 0 & 0 & 1 & 0 & 0 & 0 & 0 & 0 \\
 0 & 0 & 0 & 1 & 0 & 0 & 0 & 0 \\
 -1 & 0 & 0 & 0 & 0 & 0 & 0 & 0 \\
 0 & -1 & 0 & 0 & 0 & 0 & 0 & 0 \\
 0 & 0 & 0 & 0 & 0 & 0 & -1 & 0 \\
 0 & 0 & 0 & 0 & 0 & 0 & 0 & -1 \\
 0 & 0 & 0 & 0 & 1 & 0 & 0 & 0 \\
 0 & 0 & 0 & 0 & 0 & 1 & 0 & 0
\end{array}
\right),
\ee

\be
  \hat{{ \bm {\rm L}}}_4 = \left(
\begin{array}{cccccccc}
 0 & 0 & 0 & -1 & 0 & 0 & 0 & 0 \\
 0 & 0 & 1 & 0 & 0 & 0 & 0 & 0 \\
 0 & -1 & 0 & 0 & 0 & 0 & 0 & 0 \\
 1 & 0 & 0 & 0 & 0 & 0 & 0 & 0 \\
 0 & 0 & 0 & 0 & 0 & 0 & 0 & -1 \\
 0 & 0 & 0 & 0 & 0 & 0 & 1 & 0 \\
 0 & 0 & 0 & 0 & 0 & -1 & 0 & 0 \\
 0 & 0 & 0 & 0 & 1 & 0 & 0 & 0
\end{array}
\right),
\ee
\be
  \hat{{\bm {\rm R}}}_{\rm I} = \hat{{ \bm {\rm L}}}_{\rm I}^{t} = \hat{{ \bm {\rm L}}}_{\rm I}^{-1}.
\ee
These can be written in terms of the $SO(4)$ generators
\be\label{eq:SO4Generators}
\begin{array}{lll}
    i{\bm \alpha}^1 = i{\bm \sigma}^2 \otimes {\bm \sigma}^1~~, & i{\bm \alpha}^2 = i{\bm {\rm I}}_2 \otimes {\bm \sigma}^2~~, & i{\bm \alpha}^3 = i{\bm \sigma}^2 \otimes {\bm \sigma}^3~~, \\
     i{\bm \beta}^1 = i{\bm \sigma}^1 \otimes {\bm \sigma}^2~~, & i{\bm \beta}^2 = i{\bm \sigma}^2 \otimes {\bm {\rm I}}_2 ~~, & i{\bm \beta}^3 = i{\bm \sigma}^3 \otimes {\bm \sigma}^2
   \end{array}
\ee
as
\be\label{eq:RSSValise}
   \begin{array}{ll}
       \hat{{ \bm {\rm L}}}_1 = {\bm {\rm I}}_8 ~~, & \hat{{ \bm {\rm L}}}_2 = i {\bm {\rm I}}_2 \otimes {\bm \beta}^3 \\
       \hat{{ \bm {\rm L}}}_3 = i {\bm \sigma}^{3} \otimes {\bm \beta}^2~~, & \hat{{ \bm {\rm L}}}_4 = -i {\bm {\rm I}}_2 \otimes {\bm \beta}^1
    \end{array}
\ee
When comparing with Eqs.~(\ref{eq:cisAdinkraMatrices})and(\ref{eq:transAdinkraMatrices}), it is clear the block diagonal representation~(\ref{eq:RSSValise}) of the real scalar superfield Adinkra matrices are composed of the cis-Adinkra matrices in the upper block and the trans-Adinkra matrices in the lower block.

Furthermore, the Adinkra for these matrices is as in Fig.~\ref{fig:RSFValise}. This Adinkra is easily seen, upon comparison with Fig.~\ref{fig:VCValise}, to split into one cis- and one trans-Adinkra.  Clearly then the real scalar superfield multiplet has the SUSY enantiomer numbers $n_c = n_t = 1$.  This will be shown another way in the next section, with traces of the Adinkra matrices which is independent of node definition of the Adinkras.

\begin{figure}[!h]
   \centering
   \subfigure{ \psfrag{b1}[c][c]{\fcolorbox{black}{white}{$\int {\rm d}~dt - K'$}}
   \psfrag{b2}[c][c]{\fcolorbox{black}{white}{$2M$}}
   \psfrag{b3}[c][c]{\fcolorbox{black}{white}{$2N$}}
   \psfrag{b4}[c][c]{\fcolorbox{black}{white}{$2U_0$}}
    \psfrag{f1}[c][c]{\begin{tikzpicture}
 \node[rounded rectangle,draw,fill=white!30]{$\zeta_1' +\Lambda_2$};
\end{tikzpicture}}
   \psfrag{f2}[c][c]{\begin{tikzpicture}
 \node[rounded rectangle,draw,fill=white!30]{$\zeta_2' - \Lambda_1$};
\end{tikzpicture}}
   \psfrag{f3}[c][c]{\begin{tikzpicture}
 \node[rounded rectangle,draw,fill=white!30]{$\zeta_3' - \Lambda_4$};
\end{tikzpicture}}
   \psfrag{f4}[c][c]{\begin{tikzpicture}
 \node[rounded rectangle,draw,fill=white!30]{$-\zeta_4' - \Lambda_3$};
\end{tikzpicture}}\includegraphics[width=0.6\columnwidth]{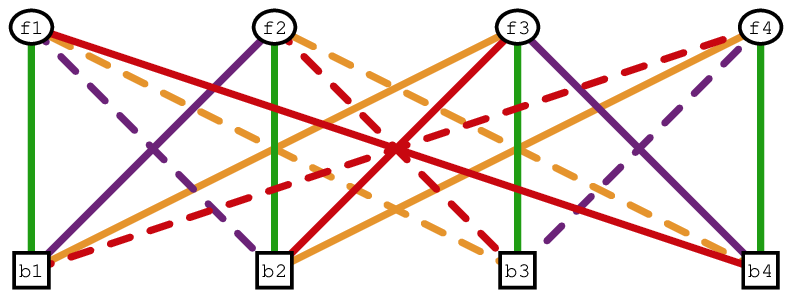}
   }
   \\
   \subfigure{
   \psfrag{b1}[c][c]{\fcolorbox{black}{white}{$-2U_2$}}
   \psfrag{b2}[c][c]{\fcolorbox{black}{white}{$\int {\rm d}~dt + K'$}}
   \psfrag{b3}[c][c]{\fcolorbox{black}{white}{$2U_1$}}
   \psfrag{b4}[c][c]{\fcolorbox{black}{white}{$-2U_3$}}
   \psfrag{f1}[c][c]{\begin{tikzpicture}
 \node[rounded rectangle,draw,fill=white!30]{ $\Lambda_1+\zeta_2'$};
\end{tikzpicture}}
   \psfrag{f2}[c][c]{\begin{tikzpicture}
 \node[rounded rectangle,draw,fill=white!30]{ $\Lambda_2-\zeta_1'$};
\end{tikzpicture}} 
    \psfrag{f3}[c][c]{\begin{tikzpicture}
 \node[rounded rectangle,draw,fill=white!30]{ $\Lambda_3-\zeta_4'$};
\end{tikzpicture}} 
    \psfrag{f4}[c][c]{\begin{tikzpicture}
 \node[rounded rectangle,draw,fill=white!30]{ $\Lambda_4+\zeta_3'$};
\end{tikzpicture}}  
   \includegraphics[width=0.6\columnwidth]{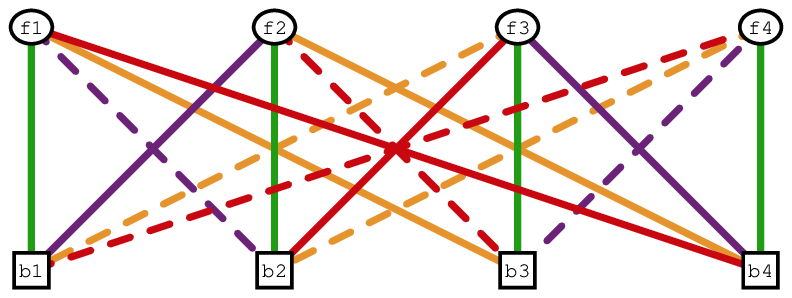}
   }
   \caption{The 8 x 8 valise Adinkra for the real scalar superfield multiplet with $n_c = n_t = 1$ (compare with Fig.~\ref{fig:VCValise}) .  All fermionic nodes have the same engineering dimension and all bosonic nodes have the same engineering dimension.} 
   \label{fig:RSFValise}
\end{figure}

\subsection{Traces}
Because the ${\bm {\rm L}}$ and $\hat{\bm {\rm L}}$ matrices are related by orthogonal transformations, the following traces
\be
Tr[{ \bm {\rm L}}_{\rm I} { \bm {\rm L}}_{\rm J}^T] = 8~ \delta_{{\rm I}{\rm J}}
\ee
\be
Tr[{ \bm {\rm L}}_{\rm I} { \bm {\rm L}}_{\rm J}^T { \bm {\rm L}}_{\rm K} { \bm {\rm L}}_{ \rm {\rm L}}^T] = 8 ~(\delta_{{\rm IJ}}\delta_{{\rm K}{ \rm L}} - \delta_{\rm IK}\delta_{{\rm J}{ \rm L}} + \delta_{{\rm I}{ \rm L}}\delta_{JK})
\ee
are identical for both sets.
We see here that the real scalar superfield multiplet fits into the conjectured trace formulas from part I~\cite{Gates:2009me}
\be\label{eq:TraceConjecture}
\eqalign{
   Tr[{\bm {\rm L}}_{\rm I} {\bm {\rm L}}_{\rm J}^t] = & 4 \, (\, n_c \,+\ n_t \,)~ \delta_{{\rm IJ}} \cr
   Tr[{\bm {\rm L}}_{\rm I} {\bm {\rm L}}_{\rm J}^t {\bm {\rm L}}_{\rm I} {\bm {\rm L}}_{\rm J}^t] = &\, 4 \, (\, n_c \,+\ n_t \,) \, (\delta_{\rm IJ}\delta_{\rm KL} - \delta_{\rm IK}\delta_{\rm JL} + \delta_{\rm IL}\delta_{\rm JK} ) \cr 
&+  4\, (\, n_c \, -\, n_t \,) \,  \e_{\rm IJKL}
}\ee
where  $n_c$ and $n_t$ are respectively, the number of cis-valise and trans-valise Adinkras contained in
an arbitrary multiplet.  We see that $n_c=n_t$ and as $n_c$ and $n_t$ are constrained by the size of the multiplet $(2d =16)$ to satisfy
\be
  4 (n_c + n_t) = d 
\ee 
we find as in Fig.~\ref{fig:RSFValise} that
\be
  n_c = n_t = 1.
\ee


\section{The 4D,\texorpdfstring{ $\cal N$}{N} = 1 Complex Linear Superfield Multiplet and Adinkras}\label{sec:CLM}

$~~~$ The 4D, $\cal N$ = 1 complex linear superfield multiplet consists of a scalar $K$, a pseudoscalar $L$, a Majorana fermion $\zeta$, a  Majorana fermion auxiliary field $\rho$, a scalar auxiliary field $M$, a pseudoscalar auxiliary field $N$,  a vector auxiliary field $V$, an axial vector auxiliary field $U$, and a Majorana fermion auxiliary field $\beta$.

\subsection{Supersymmetry Transformation Laws}  The
supersymmetry variations of the components of the complex linear superfield multiplet can be cast in the forms

\be\eqalign{
 {\rm D}_a K=& \rho_a - \zeta_a \cr
 {\rm D}_a M =&  \b_a -\frac{1}{2} (\g^\nu)_{a}^{\,\,\,d} \partial_\nu \r_d  \cr
 {\rm D}_a N =&   -i   (\g^5)_a^{\,\,\,\,d} \b_d + \frac{i}{2} (\g^5 \g^\nu)_{a}^{\,\,\,d} \partial_\nu  \r_d \cr
 {\rm D}_a L =&   i (\g^5)_a^{\,\,\,d} ( \r_d + \z_d) \cr
 {\rm D}_a U_\mu = & i (\g^5 \g_\mu)_a^{\,\,\,\,d} \b_d - i (\g^5)_a^{\,\,\,\,d}   \partial_\mu  (\r_d + 2\z_d) -  \frac{i}{2}\, (\g^5 \g^\nu \g_\mu)_{a}^{\,\,\,d} \partial_\nu(\r_d - 2\z_d) \cr
{\rm D}_a V_\mu = & - (\g_\mu)_a^{\,\,\,\,d} \b_d +   \partial_\mu (\r_a - 2\z_a) +  \frac{1}{2} ( \g^\nu \g_\mu)_{a}^{\,\,\,d} \partial_\nu (\r_d + 2\z_d) \cr
{\rm D}_a \z_b = & - i  (\g^\mu)_{ab} \partial_\mu K - (\g^5 \g^\mu)_{ab} \partial_\mu L
-  \, \frac{1}{2}( \g^5 \g^\mu)_{ab} U_\mu + i \frac{1}{2} (\g^\mu)_{ab}  V_\mu   \cr
 {\rm D}_a \r_b = &    i C_{ab} M  + (\g^5)_{a b} N + \,\frac{1}{2}( \g^5 \g^\mu)_{ab} U_\mu +  i \frac{1}{2} (\g^\mu)_{ab}  V_\mu \cr
 {\rm D}_a \b_b = & \frac{i}{2} (\g^\mu)_{ab} \partial_\mu M + \frac{1}{2} (\g^5 \g^\mu)_{ab} \partial_\mu N  + \frac{1}{2} (\g^5 \g^\mu \g^\nu)_{ab} \partial_\mu U_\nu \cr
& + \frac{1}{4} (\g^5 \g^\nu \g^\mu )_{ab}\partial_\mu U_\nu + \frac{i}{2} (\g^\mu \g^\nu)_{ab} \partial_\mu V_\nu + \frac{i}{4} \,(\g^\nu \g^\mu)_{ab} \partial_\mu V_\nu 
\cr
& + \eta^{\mu\nu}\partial_\mu \partial_\nu (- i C_{ab} K +(\g^5)_{ab} L ).
}\label{eq:DCLM1}
\ee

As in the case of the real scalar superfield multiplet, the commutator algebra for the D-operator calculated from (\ref{eq:DCLM1}) takes the form 
\begin{equation}\label{eq:Algebra}
  \{ { \rm D}_a, { \rm D}_b \} = 2 i (\gamma^\mu)_{ab} \partial_\mu
\end{equation}
for all fields in the complex linear superfield multiplet.  The D-operator (\ref{eq:DCLM1}) is an invariance, up to total derivatives, of the Lagrangian

\be\label{eq:LagrangianCLM}\eqalign{
\mathcal{L} &= -\frac{1}{2}\partial_\mu K \partial^\mu K - \frac{1}{2} \partial_\mu L \partial^\mu L - \frac{1}{2} M^2 - \frac{1}{2} N^2 +\frac{1}{4} U_\mu U^\mu  +\frac{1}{4} V_\mu V^\mu \cr
& +\frac{1}{2}i (\gamma^\mu)^{ab} \zeta_a \partial_\mu \zeta_b  + i \rho_a C^{ab}\beta_b 
}\ee
As a bridge to section~\ref{sec:CLM0}, we show the zero-brane reduction of this Lagrangian, where all fields are assumed to have only time dependence, and time derivatives are denoted by a prime $(')$:
\be\label{eq:LagrangianCLM0}\eqalign{
\mathcal{L}^{(0)} = &  \frac{1}{2}K'^2 + \frac{1}{2} L'^2  - \frac{1}{2}M^2 - \frac{1}{2}N^2 + \frac{1}{4} U_\mu U^{\mu} + \frac{1}{4} V_\mu V^\mu \cr
&+ i\frac{1}{2}(\z_1 \z_1' + \z_2 \z_2' + \z_3 \z_3' + \z_4 \z_4') + i (\r_2 \b_1 - \r_1 \b_2 + \r_3 \b_4 - \r_4 \b_3)
}\ee

This Lagrangian can be acquired by the following calculation, performed on the zero-brane:
\be
  {\rm D}_1 {\rm D}_2 {\rm D}_3 {\rm D_4} {\mathcal L}_{Superspace}^{(0)} =  {\mathcal L}^{(0)} + \mbox{total derivatives}
\ee
where ${\mathcal L}_{Superspace}^{(0)} = -\frac{1}{8}(K^2 + L^2)$, the Lagrangian for the complex linear superfield multiplet in superspace.  This calculation is shown explicitly in appendix~\ref{app:CLMConsistencyCheck}.

\subsection{Zero Brane Reduction}\label{sec:CLM0}
Before creating the Adinkra for the complex linear superfield multiplet, we must first reduce the transformation laws, Eq.~(\ref{eq:DCLM1}), to one dimension by considering the fields to have only time dependence.  The resulting SUSY transformation laws are, for the bosons:

\be
\begin{array}{llll}
{\rm D}_1 K= \r_1 - \z_1 & {\rm D}_2 K= \r_2 -\z_2 &{\rm D}_3 K=\r_3 - \z_3 & {\rm D}_4 K = \r_4 - \z_4 
\end{array}
\ee
\be
\begin{array}{ll}
{\rm D}_1 M= -\frac{1}{2} \rho'_2+\beta _1 
   & {\rm D}_2 M=\frac{\rho'_1}{2} +\beta _2 \\
{\rm D}_3 M=\frac{\rho'_4}{2} +\beta _3 
   & {\rm D}_4 M = -\frac{\rho'_3+}{2}  \beta _4
   \end{array}
\ee
\be
\begin{array}{ll}
{\rm D}_1 N=  -\frac{\rho'_3}{2} +\beta _4   
& {\rm D}_2 N= -\frac{1}{2}  \rho'_4-\beta _3 \\
{\rm D}_3 N=\frac{\rho'_1}{2}  +\beta _2 & {\rm D}_4 N = \frac{\rho'_2}{2}  -\beta _1
\end{array}
\ee
\be
\begin{array}{llll}
 {\rm D}_1 L = -\rho _4 - \z_4 & {\rm D}_2 L = \rho _3 + \z_3 &
 {\rm D}_3 L= -\rho _2 - \z_2 & {\rm D}_4 L = \rho _1 + \z_1
 \end{array}
\ee
\be
 \begin{array}{ll}
{\rm D}_1 U_0 = \beta _3+\zeta'_4+\frac{3 \rho'_4}{2} &
 {\rm D}_1 U_1 = -\beta _3-\zeta'_4+\frac{\rho'_4}{2} \\ 
{\rm D}_1 U_2 = \beta _1-\zeta'_2+\frac{\rho'_2}{2} &
 {\rm D}_1 U_3 = \beta _4-\zeta'_3+\frac{\rho'_3}{2} 
 \end{array}
 \ee
 \be
 \begin{array}{ll}
{\rm D}_2 U_0 = \beta _4-\zeta'_3-\frac{3 \rho'_3}{2} &
 {\rm D}_2 U_1= \beta _4-\zeta'_3+\frac{\rho'_3}{2} \\
{\rm D}_2 U_2 = \beta _2+\zeta'_1-\frac{\rho'_1}{2} &
 {\rm D}_2 U_3 = \beta _3+\zeta'_4-\frac{\rho'_4}{2} 
 \end{array}
 \ee
 \be
 \begin{array}{ll}
{\rm D}_3 U_0 =  -\beta _1+\zeta'_2+\frac{3 \rho'_2}{2} & {\rm D}_3 U_1 = -\beta _1+\zeta'_2-\frac{\rho'_2}{2} \\
 {\rm D}_3 U_2 = -\beta _3-\zeta'_4+\frac{\rho'_4}{2} &
  {\rm D}_3 U_3 = \beta _2+\zeta'_1-\frac{\rho'_1}{2} 
  \end{array}
  \ee
  \be
  \begin{array}{ll}
{\rm D}_4 U_0 = -\beta _2-\zeta'_1-\frac{3 \rho'_1}{2} & {\rm D}_4 U_1 = \beta _2+\zeta'_1-\frac{\rho'_1}{2} \\
{\rm D}_4 U_2 =  -\beta _4+\zeta'_3-\frac{\rho'_3}{2} & {\rm D}_4 U_3 = \beta _1-\zeta'_2+\frac{\rho'_2}{2}
   \end{array}
\ee
\be
\begin{array}{ll}
{\rm D}_1 V_0 =  \beta _2-\zeta'_1+\frac{3 \rho'_1}{2} & {\rm D}_1 V_1 = -\beta _2+\zeta'_1+\frac{\rho'_1}{2} \\
{\rm D}_1 V_2 = \beta _4+\zeta'_3+\frac{\rho'_3}{2} & {\rm D}_1 V_3 = -\beta _1-\zeta'_2-\frac{\rho'_2}{2} 
\end{array}
\ee
\be  
\begin{array}{ll} 
 {\rm D}_2 V_0 = -\beta _1-\zeta'_2+\frac{3 \rho'_2}{2} & {\rm D}_2 V_1 =-\beta _1-\zeta'_2-\frac{\rho'_2}{2} \\
 {\rm D}_2 V_2 = -\beta _3+\zeta'_4+\frac{\rho'_4}{2} & {\rm D}_2 V_3 =\beta _2-\zeta'_1-\frac{\rho'_1}{2} 
\end{array}
\ee
\be  
\begin{array}{ll}    
 {\rm D}_3 V_0 = -\beta _4-\zeta'_3+\frac{3 \rho'_3}{2} & {\rm D}_3 V_1 = -\beta _4-\zeta'_3-\frac{\rho'_3}{2} \\
 {\rm D}_3 V_2 =  -\beta _2+\zeta'_1+\frac{\rho'_1}{2} & {\rm D}_3 V_3 = -\beta _3+\zeta'_4+\frac{\rho'_4}{2} 
   \end{array}
\ee
\be  
\begin{array}{ll} 
 {\rm D}_4 V_0 = \beta _3-\zeta'_4+\frac{3 \rho'_4}{2} & {\rm D}_4 V_1 = -\beta _3+\zeta'_4+\frac{\rho'_4}{2} \\
  {\rm D}_4 V_2 = \beta _1+\zeta'_2+\frac{\rho'_2}{2} & {\rm D}_4 V_3 = \beta _4+\zeta'_3+\frac{\rho'_3}{2}
 \end{array}
\ee
and for the fermions:
\be
\begin{array}{ll}
 {\rm D}_1 \z_1 = \frac{i V_0}{2}+\frac{i V_1}{2}-i K' & {\rm D}_1 \z_2 = -\frac{i U_2}{2}-\frac{i V_3}{2} \\
  {\rm D}_1 \z_3 = \frac{i V_2}{2}-\frac{i U_3}{2} & {\rm D}_1 \z_4 = -\frac{i U_0}{2}-\frac{i U_1}{2}-i L' 
\end{array}
\ee
\be 
\begin{array}{ll}
{\rm D}_2 \z_1 = \frac{i U_2}{2}-\frac{i V_3}{2} & {\rm D}_2 \z_2 =\frac{i V_0}{2}-\frac{i V_1}{2}-i K' \\
 {\rm D}_2 \z_3 = \frac{i U_0}{2}-\frac{i U_1}{2}+i L' & {\rm D}_2 \z_4 = \frac{i U_3}{2}+\frac{i V_2}{2} 
\end{array}
\ee
\be 
\begin{array}{ll}
{\rm D}_3 \z_1 = \frac{i U_3}{2}+\frac{i V_2}{2} & {\rm D}_3 \z_2 = -\frac{i U_0}{2}+\frac{i U_1}{2}-i L' \\
 {\rm D}_3 \z_3 = \frac{i V_0}{2}-\frac{i V_1}{2}-i K' & {\rm D}_3 \z_4 = \frac{i V_3}{2}-\frac{i U_2}{2} 
\end{array}
\ee
\be 
\begin{array}{ll}
{\rm D}_4 \z_1 = \frac{i U_0}{2}+\frac{i U_1}{2}+i L' & {\rm D}_4 \z_2 = \frac{i V_2}{2}-\frac{i U_3}{2} \\
 {\rm D}_4 \z_3 = \frac{i U_2}{2}+\frac{i V_3}{2} & {\rm D}_4 \z_4 =\frac{i V_0}{2}+\frac{i V_1}{2}-i K' 
\end{array}
\ee
\be
\begin{array}{ll}
{\rm D}_1 \r_1 = \frac{i V_0}{2}+\frac{i V_1}{2} & {\rm D}_1 \r_2 = -i M+\frac{i U_2}{2}-\frac{i V_3}{2} \\
 {\rm D}_1 \r_3 = -i N+\frac{i U_3}{2}+\frac{i V_2}{2} & {\rm D}_1 \r_4 = \frac{i U_0}{2}+\frac{i U_1}{2} 
\end{array}
\ee
\be
\begin{array}{ll}
{\rm D}_2 \r_1 = i M-\frac{i U_2}{2}-\frac{i V_3}{2} & {\rm D}_2 \r_2 = \frac{i V_0}{2}-\frac{i V_1}{2} \\
{\rm D}_2 \r_3 = \frac{i U_1}{2}-\frac{i U_0}{2} & {\rm D}_2 \r_4 = -i N-\frac{i U_3}{2}+\frac{i V_2}{2} 
\end{array}
\ee
\be
\begin{array}{ll}
{\rm D}_3 \r_1 =  i N-\frac{i U_3}{2}+\frac{i V_2}{2} & {\rm D}_3 \r_2 =  \frac{i U_0}{2}-\frac{i U_1}{2} \\
 {\rm D}_3 \r_3 =  \frac{i V_0}{2}-\frac{i V_1}{2} & {\rm D}_3 \r_4 = i M+\frac{i U_2}{2}+\frac{i V_3}{2} 
 \end{array}
\ee
\be
\begin{array}{ll}
{\rm D}_4 \r_1 =  -\frac{i U_0}{2}-\frac{i U_1}{2} & {\rm D}_4 \r_2 =  i N+\frac{i U_3}{2}+\frac{i V_2}{2} \\
 {\rm D}_4 \r_3 =  -i M-\frac{i U_2}{2}+\frac{i V_3}{2} & {\rm D}_4 \r_4 = \frac{i V_0}{2}+\frac{i V_1}{2}
\end{array}
\ee
\be
\begin{array}{ll}
{\rm D}_1 \b_1=   \frac{i U'_2}{4} -\frac{i
   V'_3}{4} +\frac{i M'}{2} &
{\rm D}_1 \b_2=  \frac{3i V'_0}{4} -\frac{i V'_1}{4}  - i K'' \\
{\rm D}_1 \b_3=  \frac{3i U'_0}{4} -\frac{i 
   U'_1}{4}  + i L'' &
{\rm D}_1 \b_4=    \frac{i U'_3}{4} +\frac{i V'_2}{4} +\frac{iN'}{2}  \end{array}
\ee
\be
\begin{array}{ll}
{\rm D}_2 \b_1=   -\frac{3i  V'_0}{4} -\frac{i V'_1}{4}   + i K'' &
{\rm D}_2 \b_2=   \frac{i U'_2}{4} +\frac{i    V'_3}{4} +\frac{i}{2} M' \\
{\rm D}_2 \b_3= \frac{i U'_3}{4} -\frac{i 
   V'_2}{4} -\frac{i N'}{2}  &
{\rm D}_2 \b_4=  \frac{3i U'_0}{4} +\frac{i U'_1}{4}  + i L''  \end{array}
\ee
\be
\begin{array}{ll}
{\rm D}_3 \b_1=  -\frac{i U'_1}{4} -\frac{3i 
   U'_0}{4}   - i L''  &
{\rm D}_3 \b_2=   \frac{i U'_3}{4} -\frac{i    V'_2}{4} +\frac{iN' }{2} \\
{\rm D}_3 \b_3=   -\frac{i U'_2}{4} -\frac{i 
   V'_3}{4} +\frac{iM'}{2}  &
{\rm D}_3 \b_4=    -\frac{3i V'_0}{4} -\frac{i V'_1}{4}  + i K''\end{array}
\ee
\be
\begin{array}{ll}
{\rm D}_4 \b_1=   \frac{i U'_3}{4} +\frac{i 
   V'_2}{4} -\frac{iN'}{2}   &
{\rm D}_4 \b_2 =  -\frac{3i U'_0}{4} +\frac{i 
   U'_1}{4}  - i L'' \\
{\rm D}_4 \b_3=  \frac{3i  V'_0}{4} -\frac{i  V'_1}{4}   - i K'' &
{\rm D}_4 \b_4=  -\frac{i U'_2}{4} +\frac{i 
   V'_3}{4} +\frac{i M'}{2} 
\end{array}
\ee
With the rules for Adinkras reviewed in appendix~\ref{app:CM}, it can be seen that these transformation laws can be succinctly displayed as the Adinkra in Fig.~\ref{fig:12by12Adinkra}.

\begin{figure}[!htbp]
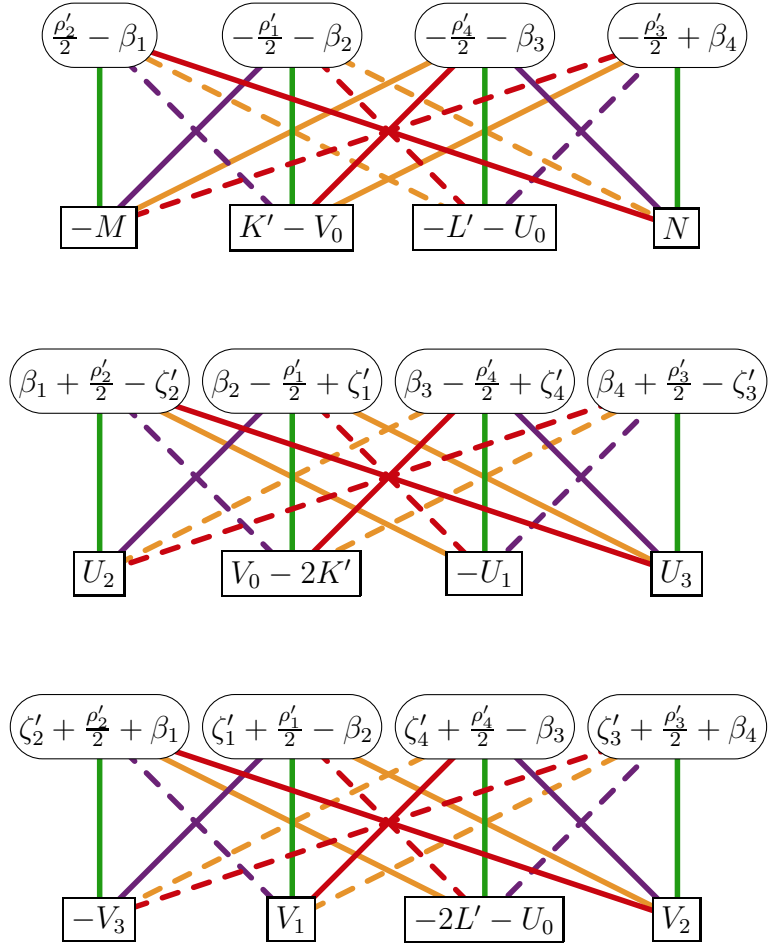

\centering
\subfigure{
\psfrag{b1}[c][c]{\fcolorbox{black}{white}{$-M$}}
\psfrag{b2}[c][c]{\fcolorbox{black}{white}{$K'-V_0$}}
\psfrag{b3}[c][c]{\fcolorbox{black}{white}{$-L'-U_0$}}
\psfrag{b4}[c][c]{\fcolorbox{black}{white}{$N$}}
\psfrag{f1}[c][c]{\begin{tikzpicture} 
\node[rounded rectangle, inner xsep = 0pt ,draw,fill=white!30]{$\frac{\rho'_2}{2}-\beta_1$};
\end{tikzpicture}}
\psfrag{f2}[c][c]{\begin{tikzpicture}
\node[rounded rectangle, inner xsep = 0pt, draw,fill=white!30]{$-\frac{\rho'_1}{2}-\beta_2$};
\end{tikzpicture}}
\psfrag{f3}[c][c]{\begin{tikzpicture}
 \node[rounded rectangle, inner xsep = 0pt ,draw,fill=white!30]{$-\frac{\rho'_4}{2}-\beta_3$};
\end{tikzpicture}}
\psfrag{f4}[c][c]{\begin{tikzpicture}
 \node[rounded rectangle, inner xsep = 0pt ,draw,fill=white!30]{$-\frac{\rho'_3}{2}+\beta_4$};
\end{tikzpicture}}
\includegraphics[width = 0.6\columnwidth]{ChiralValisewFields}
}
\quad
\subfigure{
\psfrag{b1}[c][c]{\fcolorbox{black}{white}{$U_2$}}
\psfrag{b2}[c][c]{\fcolorbox{black}{white}{$V_0-2K'$}}
\psfrag{b3}[c][c]{\fcolorbox{black}{white}{$-U_1$}}
\psfrag{b4}[c][c]{\fcolorbox{black}{white}{$U_3$}}
\psfrag{f1}[c][c]{\begin{tikzpicture}
 \node[rounded rectangle, inner xsep = 0pt ,draw,fill=white!30]{$\beta_1 + \frac{\rho'_2}{2} - \zeta'_2$};
\end{tikzpicture}}
\psfrag{f2}[c][c]{\begin{tikzpicture}
 \node[rounded rectangle, inner xsep = 0pt ,draw,fill=white!30]{$\beta_2 - \frac{\rho'_1}{2} + \zeta'_1$};
\end{tikzpicture}}
\psfrag{f3}[c][c]{\begin{tikzpicture}
 \node[rounded rectangle, inner xsep = 0pt ,draw,fill=white!30]{$\beta_3 - \frac{\rho'_4}{2} + \zeta'_4$};
\end{tikzpicture}}
\psfrag{f4}[c][c]{\begin{tikzpicture}
 \node[rounded rectangle, inner xsep = 0pt ,draw,fill=white!30]{$\beta_4 + \frac{\rho'_3}{2} - \zeta'_3$};
\end{tikzpicture}}
\includegraphics[width=0.6\columnwidth]{VectorValisewFields}
}
\quad
\subfigure{
\psfrag{f1}[c][c]{\begin{tikzpicture}
 \node[rounded rectangle, inner xsep = 0pt, draw,fill=white!30]{$\zeta'_2+\frac{\rho'_2}{2} +\beta_1$};
\end{tikzpicture}}
\psfrag{f2}[c][c]{\begin{tikzpicture}
 \node[rounded rectangle, inner xsep = 0 pt, draw,fill=white!30]{$\zeta'_1+\frac{\rho'_1}{2} -\beta_2$};
\end{tikzpicture}}
\psfrag{f3}[c][c]{\begin{tikzpicture}
 \node[rounded rectangle, inner xsep = 0 pt, draw,fill=white!30]{$\zeta'_4+\frac{\rho'_4}{2} -\beta_3$};
\end{tikzpicture}}
\psfrag{f4}[c][c]{\begin{tikzpicture}
 \node[rounded rectangle, inner xsep = 0 pt, draw,fill=white!30]{$\zeta'_3+\frac{\rho'_3}{2} +\beta_4$};
\end{tikzpicture}}
\psfrag{b1}[c][c]{\fcolorbox{black}{white}{$-V_3$}}
\psfrag{b2}[c][c]{\fcolorbox{black}{white}{$V_1$}}
\psfrag{b3}[c][c]{\fcolorbox{black}{white}{$-2L'-U_0$}}
\psfrag{b4}[c][c]{\fcolorbox{black}{white}{$V_2$}}
\includegraphics[width=0.6\columnwidth]{VectorValisewFields}
}
\caption{The $12 \times 12$ valise Adinkra for the complex linear superfield multiplet with $n_c = 1$ and $n_t = 2$ (compare with Fig.~\ref{fig:VCValise}). All fermionic nodes have the same engineering dimension and all bosonic nodes have the same engineering dimension.}
\label{fig:12by12Adinkra}
\end{figure}

Next, we define fields from the nodes of Fig.~\ref{fig:12by12Adinkra}
\be
  \Phi =
\left(
\begin{array}{c}
 -M \\
 K'-V_0 \\
 -L'-U_0 \\
 N \\
 U_2 \\
 V_0-2 K' \\
 -U_1 \\
 U_3 \\
 -V_3 \\
 V_1 \\
 -2 L'-U_0 \\
 V_2
\end{array}
\right)
,~~~i \Psi = \left(
\begin{array}{c}
 \frac{\r'_2}{2}-\beta _1 \\
 -\beta _2-\frac{\r'_1}{2} \\
 -\beta _3-\frac{\r'_4}{2} \\
 \beta _4-\frac{\r'_3}{2} \\
 \beta _1-\z'_2+\frac{\r'_2}{2} \\
 \beta _2+\z'_1-\frac{\r'_1}{2} \\
 \beta _3+\z'_4-\frac{\r'_4}{2} \\
 \beta _4-\z'_3+\frac{\r'_3}{2} \\
 \beta _1+\z'_2+\frac{\r'_2}{2} \\
 -\beta _2+\z'_1+\frac{\r'_1}{2} \\
 -\beta _3+\z'_4+\frac{\r'_4}{2} \\
 \beta _4+\z'_3+\frac{\r'_3}{2}
\end{array}
\right)
\ee
such that
\be
  {\rm D}_{\rm I} \Phi_{\hat{i}} = i \left( {\rm L }_{\rm I} \right)_{i\hat{j}}\Psi_{\hat{j}},~~~{\rm D}_{\rm I} \Psi_{\hat{k}} = \left( {\rm R }_{\rm I} \right)_{\hat{k}i}\Phi_i'.
\ee
where the ${\bm {\rm L}}$ and ${\bm {\rm R}}$ Adinkra matrices are
\be\label{eq:CLSValise}
\begin{array}{llll}
{\bm {\rm L}}_1 =  {\bm {\rm  I}}_3  \otimes {\bm {\rm I}}_4 & ,& {\bm {\rm L}}_2 = i {\bm {\rm I}}_3 \otimes {\bm {\rm \beta}}_3 &, \\
{\bm {\rm L}}_3 = i \left(\begin{array}{ccc}
                        1 & 0 & 0 \\
                        0 & -1 & 0 \\
                        0 & 0 & -1
                   \end{array}\right) \otimes  {\bm {\rm \beta}}_2&, &{\bm {\rm L}}_4 = - i {\bm {\rm I}}_3 \otimes {\bm {\rm \beta}}_1 & ,
\end{array}
\ee
and
\be
  {\bm {\rm R}}_{\rm I} = {\bm {\rm L}}_{\rm I}^t = {\bm {\rm L}}_{\rm I}^{-1}~~~.
\ee
The matrices in Eq.~(\ref{eq:CLSValise}) are written in terms of the $SO(4)$ generators in Eq.~(\ref{eq:SO4Generators}) and comparing with Eqs.~(\ref{eq:cisAdinkraMatrices})and(\ref{eq:transAdinkraMatrices}), it is clear that the block diagonal representation~(\ref{eq:CLSValise}) of the complex linear superfield Adinkra matrices are composed of the cis-Adinkra matrices in the upper block and trans-Adinkra matrices in each of the lower two blocks. It is then clear from Eq.~(\ref{eq:CLSValise}) that the complex linear superfield has the SUSY enantiomer numbers $n_c = 1, n_t = 2$. This same conclusion can be easily arrived at from comparing Fig.~\ref{fig:12by12Adinkra} to the definitions of the cis- and trans-valise Adinkras in Fig. \ref{fig:VCValise}. We will confirm the SUSY enantiomer numbers through basis independent traces in section~\ref{sec:CLSTraces}. Furthermore, making the substitution ${\bm \beta}_2 \to -{\bm \beta}_2$ would transform the Adinkra matrices in Eq.~\ref{eq:CLSValise} into the $n_c = 2, n_t = 1$ SUSY representation.

We can write the matrices in Eq.~(\ref{eq:CLSValise}) as
\be\label{eq:CLSValise12x12}
\begin{array}{llll}
{\bm {\rm L}}_1 =  {\bm {\rm  I}}_3  \otimes {\bm {\rm I}}_4 & ,& {\bm {\rm L}}_2 = i {\bm {\rm I}}_3 \otimes {\bm {\rm \beta}}_3 &, \\
{\bm {\rm L}}_3 = i \left({\bm {\rm h}}_4 - \frac{1}{3} {\bm {\rm I}}_3 + \frac{1}{3} {\bm {\rm h}}_5\right)\otimes {\bm {\rm \beta}}_2 &, &{\bm {\rm L}}_4 = - i {\bm {\rm I}}_3 \otimes {\bm {\rm \beta}}_1 & ,
\end{array}
\ee
where they are written in terms of the in the following $ 12 \times 12$ basis
\be \eqalign{
{\bm {\cal M}}_{{12} \times {12}} ~&=~ a_{0} \,  {\bm {\rm I}}_3\otimes {\bm {\rm I}}_4 
 ~+~ \sum_I \, a_{I} \,   {\bm {\rm a}}{}_I  \otimes {\bm {\rm I}}_4   ~+~  \sum_{\D} 
b_{\D} \,   {\bm {\rm h}}{}_{\D}  \otimes {\bm {\rm I}}_4   \cr
~&~~~~+~ \sum_I c_I   {\bm {\rm I}}_3 \otimes {\bm {\a}}_I 
 ~+~ \sum_{I , \, K} \, d_{I \, K} \,    
  {\bm {\rm a}}{}_I   \otimes {\bm {\a}}_K  
~+~   \sum_{\D , \, I} \, e_{\D \, I} \, 
 \,  {\bm {\rm h}}{}_{\D}   \otimes {\bm {\a}}_I      \cr
~&~~~~+~ \sum_I {\Tilde c}_I   {\bm {\rm I}}_3 \otimes {\bm {\b}}_I 
 ~+~ \sum_{I , \, K} \, {\Tilde d}_{I \, K} \,    
  {\bm {\rm a}}{}_I \otimes {\bm {\b}}_K    ~+~   \sum_{\D , \, I} \, {\Tilde e}_{\D \, I} \, 
 \,     {\bm {\rm h}}{}_{\D}  \otimes {\bm {\b}}_I    \cr
~&~~~~+~ \sum_{I, \, J} f_{I , \,J}   {\bm {\rm I}}_3 \otimes  {\bm {\a}}_I  {\bm {\b}}_J 
 ~+~ \sum_{I , \, J, \, K} \, g_{I, \, J, \, K} \,    
  {\bm {\rm a}}{}_I \otimes {\bm {\a}}_J   {\bm {\b}}_K    \cr
~&~~~~+~   \sum_{\D , \, I , \, J } \, h_{\D , \, I \, J} \, 
 \,     {\bm {\rm h}}{}_{\D}  \otimes {\bm {\a}}_I   {\bm {\b}}_J    ~~~,
}\ee
where ${\bm {\rm a}}_I$ and ${\bm {\rm h}}_{\D}$ are the $SL(3)$ generators
\be\label{eq:sl3a}\eqalign{
{\bm {\rm a}}_1 =  \left(
\begin{array}{ccc}
 0 & 0 & 0 \\
 0 & 0 & -1 \\
 0 & 1 & 0
\end{array}
\right)&,~~~{\bm {\rm a}}_2 =\left(
\begin{array}{ccc}
 0 & 0 & 1 \\
 0 & 0 & 0 \\
 -1 & 0 & 0
\end{array}
\right), \cr
{\bm {\rm a}}_3 =& \left(
\begin{array}{ccc}
 0 & -1 & 0 \\
 1 & 0 & 0 \\
 0 & 0 & 0
\end{array}
\right),
}\ee
and
\be\label{eq:sl3h}\eqalign{
{\bm {\rm h}}_1 = \left(
\begin{array}{ccc}
 0 & 0 & 0 \\
 0 & 0 & 1 \\
 0 & 1 & 0
\end{array}
\right)&,~~~{\bm {\rm h}}_2 = \left(
\begin{array}{ccc}
 0 & 0 & 1 \\
 0 & 0 & 0 \\
 1 & 0 & 0
\end{array}
\right),\cr
{\bm {\rm h}}_3 = & \left(
\begin{array}{ccc}
 0 & 1 & 0 \\
 1 & 0 & 0 \\
 0 & 0 & 0
\end{array}
\right),\cr
{\bm {\rm h}}_4 = \left(
\begin{array}{ccc}
 1 & 0 & 0 \\
 0 & -1 & 0 \\
 0 & 0 & 0
\end{array}
\right)&,~~~{\bm {\rm h}}_5 = \left(
\begin{array}{ccc}
 1 & 0 & 0 \\
 0 & 1 & 0 \\
 0 & 0 & -2
\end{array}
\right)
}\ee

\subsection{Traces} \label{sec:CLSTraces}
Calculating the traces
\be
\begin{array}{ll}
Tr[{\bm {\rm L}}_{\rm I} ({\bm {\rm L}}_{\rm J})^{t}] = 12~\delta_{\rm IJ}
\end{array}
\ee
and
\be
  \begin{array}{l}
 Tr[{\bm {\rm L}}_{\rm I} ({\bm {\rm L}}_{\rm J})^t {\bm {\rm L}}_{\rm K} ({\bm {\rm L}}_{\rm I})^t] = 12 (\, \delta_{\rm IJ}\delta_{\rm KL} - \, \delta_{\rm IK}\delta_{\rm JL} +  \,\delta_{\rm IL}\delta_{\rm JK} )
 -4 \,\,\epsilon_{\rm IJKL}
 \end{array}
\ee 
and comparing with the conjecture in Eq.~(\ref{eq:TraceConjecture}), for  the complex linear superfield multiplet we find
\be
  n_c = 1 ~~~, ~~~  n_t = 2  ~~~.
\ee
This is a basis independent confirmation of our findings from section~\ref{sec:CLM0}, i.e., orthogonal transformations on the fields like those performed on the real scalar superfield multiplet in section~\ref{sec:RSSO8Transforms} won't change this result.

\section{Characteristic Polynomials for Adinkras}\label{sec:CharPs}
In this final section, we investigate yet another way to distinguish SUSY representations, the characteristic polynomial
\be
   P_{\rm I}({\mathcal J}) \equiv \det \left(
       \begin{array}{cc}
              {\mathcal J} {\bm {\rm I}}_n & - {\bm {\rm L}}_{\rm I} \\
              - {\bm {\rm R}}_{\rm I} & {\mathcal J} {\bm {\rm I}}_m
       \end{array}
       \right) ,~~~\mbox{no $I$ sum}
\ee
for $n \times m$ Adinkra matrices ${\bm {\rm L}}_{\rm I}$ and $m \times n$ Adinkra matrices ${\bm {\rm R}}_{\rm I}$ with ${\mathcal J}$ an arbitrary constant.  Results for several $4D$, ${\mathcal N} = 1$ systems are listed in Tab.~\ref{tab:CharPs}, which are all color independent.

\begin{table}[!htbp]
\centering
   \begin{tabular}{c|c|c|c}
         SUSY rep. &  $P_{\rm I}({\mathcal J})$ & $d_L$ & $d_R$ \\
         \hline
         off-shell chiral multiplet (case $I$) & $({\mathcal J}^2 - 1)^4$ & 4 & 4  \\
         \hline
         on-shell chiral multiplet (case $II$) & ${\mathcal J}^2 ({\mathcal J}^2 - 1)^2$ & 2 & 4 \\
		 \hline
		 off-shell tensor multiplet (case $III$) & $({\mathcal J}^2 - 1)^4$  & 4 & 4 \\ 
		 \hline
         double tensor multiplet (case $IV$)  & ${\mathcal J}^2 ({\mathcal J}^4-3{\mathcal J}^2 +2)^2$ & 6 & 4\\ 
         \hline
         off-shell vector multiplet (case $V$)  & $({\mathcal J}^2 - 1)^4$  & 4 & 4 \\  
         \hline
         on-shell vector multiplet (case $VI$) & ${\mathcal J} ({\mathcal J}^2 - 1)^3$ & 3 & 4\\
         \hline
         real scalar superfield multiplet & $({\mathcal J}^2 - 1)^8$  & 8 & 8\\
         \hline
         complex linear superfield multiplet & $({\mathcal J}^2 - 1)^{12}$ & 12 & 12                                      
   \end{tabular}
   \caption{Characteristic polynomials for Adinkras of several $4D$, ${\mathcal N}=1$ systems with $d_L$ ($d_R$) bosonic (fermionic) nodes. The Adinkra matrices for cases $I$ through $VI$ are as in Part I~\cite{Gates:2009me}.}
   \label{tab:CharPs}
\end{table}

The off-shell cases $I$, $III$, $V$, the real scalar, and complex linear superfield multiplets clearly follow a similar pattern here.  This stems from the fact that each of their Adinkras have equal numbers of bosonic and fermionic nodes and as a consequence have square ${\bm {\rm L}}_{\rm I}$ and ${\bm {\rm R}}_{\rm I}$ matrices.   For mutually commuting $d \times d$ matrices ${\bm{\rm A}}$, ${\bm{\rm B}}$, ${\bm{\rm C}}$, and ${\bm{\rm D}}$ we have the mathematical identity~\cite{Silvester:2000mg}
\be
   \det \left( \begin{array}{cc}
              {\bm{\rm A}} & {\bm {\rm B}} \\
              {\bm {\rm C}} & {\bm {\rm D}}
       \end{array}
       \right) = \det\left( {\bm{\rm A}} {\bm {\rm D}} - {\bm {\rm B}}{\bm {\rm C}} \right) .
\ee
This leads us to the following result for the square Adinkra matrices
\be\eqalign{
   P_{\rm I}({\mathcal J}) \equiv &\det \left(
       \begin{array}{cc}
              {\mathcal J} {\bm {\rm I}}_d & - {\bm {\rm L}}_{\rm I} \\
              - {\bm {\rm R}}_{\rm I} & {\mathcal J} {\bm {\rm I}}_d
       \end{array}
       \right)  \cr
       = &\det \left( {\mathcal J}^2 {\bm {\rm I}}_d - {\bm {\rm L}}_{\rm I} {\bm {\rm R}}_{\rm I}  \right)        
},~~~\mbox{no $I$ sum}\ee
which, owing to the orthogonality of these off-shell $d \times d$ Adinkra matrices ${\bm {\rm R}}_{\rm I} = {\bm {\rm L}}_{\rm I}^{-1} = {\bm {\rm L}}_{\rm I}^{t} $, leads us to the formula
\be
   P_{\rm I}({\mathcal J}) = ({\mathcal J}^2 - 1)^d,~~~d_L = d_R = d.
\ee
Furthermore, this result is independent of the Adinkraic basis, as $O(d)$ transformations, e.g. Eqs.~(\ref{eq:calX8}) and (\ref{eq:calY8}), will clearly not change the result.  Cases $II$, $IV$, and $VI$ clearly do not fit so nicely into such a succinct formula, as $d_L \ne d_R$ in each of these cases. It is interesting to note that each of the characteristic polynomials in Tab.~\ref{tab:CharPs} do follow the pattern
\be
   P_{\rm I}({\mathcal J}) = {\mathcal J}^{|d_L - d_R|} \times (\mbox{Polynomial in}~{\mathcal J}).
\ee 
\noindent This leads us to the conjecture:

 \vspace{.05in}
 \begin{center}
 \parbox{4in}{\it The number of factors of ${\mathcal J}$ that factor out of the characteristic polynomial is equal to the absolute value of the difference in bosonic and fermionic nodes. }
 \end{center}  
 
\noindent It is not clear at present what information the left over polynomial multiplying the $|d_L - d_R|$ factors of ${\mathcal J}$ conveys.

Finally, we point out that case $IV$ is unique.  As shown in Part I~\cite{Gates:2009me}, its algebra does not close and is an on-shell representation with no known off-shell representation.  Along this line of thought, we point out similarities between cases $IV$ and $II$. These have the same value of $|d_L - d_R|$ and so their characteristic polynomials follow the same pattern: $P_{\rm I}({\mathcal J})={\mathcal J}^2 \times (\mbox{Polynomial in}~{\mathcal J})^2$.  This is not too surprising as in~\cite{Gates:2009me} transformations relating these two multiplets were pointed out, though an exact mapping was not found.    In fact these systems have been known to be related for quite some time as the double tensor multiplet was spawned in~\cite{Freedman:1977pa} over the relationship between antisymmetric tensor fields and scalar particles of supergravity.  

\section{Conclusion}
Through the genomics project, we seek to find a categorization system for SUSY representations in analogy to the representation theory of Lie algebras.  In part I~\cite{Gates:2009me}, we found two distinct Adinkras, cis- and trans-, that we gave in the present work the SUSY enantiomer numbers $n_c=1,~n_t=0$ and $n_c=0,~n_t=1$, respectively.  In part I~\cite{Gates:2009me}, it was conjectured that these cis- and trans-Adinkras were a basis from which all off-shell $4D$, ${\mathcal N}=1$ SUSY representations could be described.  In our work here in part II, we found this to be the case for two more off-shell $4D$, ${\mathcal N}=1$ SUSY representations.  These two systems are the real scalar ($n_c = n_t = 1$) and complex linear ($n_c = 1,~n_t = 2$) superfield multiplets.  

We showed two methods of finding these SUSY enantiomer numbers, one graphical, another computational.  The graphical way is of course to decompose the Adinkra into its cis- and trans- components.  The computational method was using traces of the Adinkra matrices, termed chromocharacters in the previous work~\cite{Gates:2009me}.  Both methods arrived at the same numbers, as expected.

Finally, in our continuing quest to find various mathematical and graphical objects with which to classify SUSY representations, we introduced the SUSY characteristic polynomial for Adinkras.  In the eight representations investigated, we saw the consistent result that the polynomial encodes the precise numerical discrepancy in numbers of bosonic and fermionic nodes in the Adinkra.  Whether this trend is a feature of all Adinkras or not, and what other information the polynomial holds, both remain to be seen.  We will continue investigations of polynomials such as these in future works, as well as derive the Adinkras and SUSY enantiomer numbers of other systems to continue building the SUSY representation table. 
   
\vspace{.05in}
 \begin{center}
 \parbox{4in}{{\it ``Details create the big picture.''}\,\,-\,\, Sanford I. Weill}
 \end{center}

\section*{Acknowledgements}
This research was supported in part by the endowment of the John S.~Toll Professorship, the University of Maryland 
Center for String \& Particle Theory, National Science Foundation Grant PHY-0354401.  SJG  and KS offer additional 
gratitude to the M.\ L.\ K. Visiting Professorship and to the M.\ I.\ T.\ Center for Theoretical Physics for support and hospitality facilitating part of this work.
We thank Leo Rodriguez, Abdul Khan, and the many other students who toiled over various aspects of the elusive complex linear superfield multiplet Adinkra.  Many of these students were a part of various SSTPRS summer school sessions held either at the University of Maryland or the University of Iowa, to which we also extend gratitude.  We thank Tristan H\"ubsch for recently renewing our interest in this difficult problem.   Most Adinkras were drawn with the aid of \emph{Adinkramat} \copyright ~2008 by G. Landweber.

\appendix
\section{Adinkra Review of Genomics I}\label{app:AdinkraReview}
In this appendix we review the construction of the Adinkras for the $4D$ $\cal N$ = 1 chiral and vector multiplets found in ~\cite{Gates:2009me}, where the SUSY transformation laws for these systems were studied in depth.  Here, we review their zero-brane reduction and how to construct the Adinkras from these transformation laws.  Also, it is reviewed that the chiral multiplet and vector multiplet have SUSY enantiomer numbers ($n_c=1, n_t = 0$) and $(n_c=0, n_t = 1)$.  The chiral multiplet valise Adinkra is then precisely the cis-Adinkra and the vector multiplet is precisely the trans-Adinkra.  These two Adinkras are shown in the body of this paper to be the building blocks of both the real scalar and complex linear superfield multiplets valise Adinkras.

\subsection{The \texorpdfstring{$4D$}{4D} \texorpdfstring{$\cal N$ =1}{N=1} Chiral Multiplet}\label{app:CM}
The zero-brane reduced SUSY transformation laws for the chiral multiplet, which are a symmetry of the Lagrangian~(\ref{eq:CMLagrangian0}), are:
\be
\begin{array}{llll}
{\rm D}_1 A = \, \psi_1 & {\rm D}_2 A = \, \psi_2 & {\rm D}_3A=\, \psi_3  &  
{\rm D}_4 A=\, \psi_4\\
{\rm D}_1 B = -  \, \psi_4  & {\rm D}_2B= \, \psi_3 &  {\rm D}_3B=- \, \psi_2 &  
{\rm D}_4B=\, \psi_1 \\
{\rm D}_1 F =   \psi_2'  & {\rm D}_2F=-  \psi_1' &  {\rm D}_3F= - \psi_4' 
&  {\rm D}_4F=  \psi_3'  \\
{\rm D}_1 G = -  \psi_3' & {\rm D}_2G= -\psi_4' &  {\rm D}_3G=  \psi_1'  
 &  {\rm D}_4G=  \psi_2'  .
\end{array}
   \label{eq:chiD0C}
\ee
and

\be
\begin{array}{llll}
{\rm D}_1 \psi_1 =  i A' & {\rm D}_2 \psi_1 = -i F &  {\rm D}_3\psi_1= i G  &  
{\rm D}_4\psi_1= i B' \\
{\rm D}_1 \psi_2 = i F  & {\rm D}_2\psi_2= i A' &  {\rm D}_3\psi_2=-i   
B' &  {\rm D}_4\psi_2= i G \\
{\rm D}_1 \psi_3 = - i G  & {\rm D}_2\psi_3=  B' &{\rm D}_3\psi_3= i A'
 &  {\rm D}_4\psi_3=  i F \\
{\rm D}_1 \psi_4 = -i B' & {\rm D}_2\psi_4=- i G & {\rm D}_3\psi_4= - i F  
 &  {\rm D}_4\psi_4= i A'  .
\end{array}
  \label{eq:chiD0H}
\ee
where a prime $(')$ denotes a time derivative.

Now we show how to build an Adinkra from these transformation laws.  Adinkras are defined by fermionic and bosonic nodes and colored connections as in fig.~\ref{fig:AdinkraRules}.  The rules are as follows:

\begin{enumerate}
\item The color encodes the identity of the transformation law.  Here and throughout this paper, we hold to the conventions of~\cite{Gates:2009me} and identify the colors as \textcolor{AdinkraGreen}{${\rm D}_1$}, \textcolor{AdinkraViolet}{${\rm D}_2$}, \textcolor{AdinkraOrange}{${\rm D}_3$}, and  \textcolor{AdinkraRed}{${\rm D}_4$}.  
\item Applying the ${\rm D}$-operator to the lower node yields the upper node.
\item Applying the ${\rm D}$-operator to the upper node yields $\partial_t$ acting on the lower node.
\item A dashed line encodes an overall additional minus sign in both associated transformations. 
\item Transformations from fermion to boson yield an additional factor of $i$ multiplying the boson.
\item Field nodes on the same horizontal level have the same engineering dimension.  Engineering dimensions increase by one half when ascending to the row directly above.
\end{enumerate}

\begin{figure}[!h]
\centering
\subfigure[]{\psfrag{s1}[c][c]{
\begin{tikzpicture}
 \node[rounded rectangle,draw,fill=white!30]{$\psi_1$};
 \end{tikzpicture}~}
\psfrag{A}[c][c]{\fcolorbox{black}{white}{$A$}}
\psfrag{equalto}[c][c]{$=\left\{\begin{array}{l}
                             \mbox{\textcolor{AdinkraGreen}{${\rm D}_1$}} \psi_1 = i A' \\
                             \mbox{\textcolor{AdinkraGreen}{${\rm D}_1$}} A = \psi_1
                             \end{array}
                              \right. 
                             $}
                             \label{fig:ReadingAnAdinkraGreen1}\includegraphics[width=0.3\columnwidth]{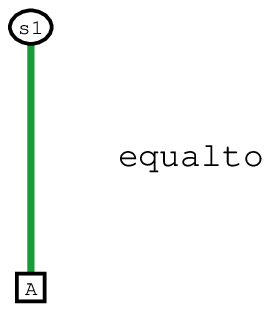}}
\quad\quad
\subfigure[]{
\psfrag{B}[c][c]{\fcolorbox{black}{white}{$B$}}
\psfrag{s2}[c][c]{
\begin{tikzpicture}
 \node[rounded rectangle,draw,fill=white!30]{$\psi_2$};
 \end{tikzpicture}~}
\psfrag{equalto}[c][c]{$=\left\{\begin{array}{l}
                             \mbox{\textcolor{AdinkraOrange}{${\rm D}_3$}} \psi_2 = -i B' \\
                             \mbox{\textcolor{AdinkraOrange}{${\rm D}_3$}} B = -\psi_2
                             \end{array}
                              \right. 
                             $}\label{fig:ReadingAnAdinkraOrange1}\includegraphics[width=0.3\columnwidth]{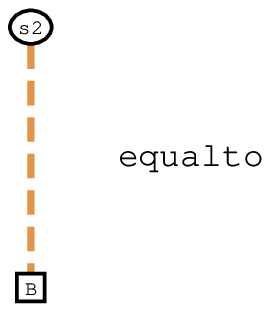}}
\caption{How to read SUSY tranformation laws from an Adinkra.}
\label{fig:AdinkraRules}  
\end{figure}

With these simple rules, we can easily see that the transformation laws~(\ref{eq:chiD0C})and~(\ref{eq:chiD0H}) encode the cis-Adinkra in Fig.~\ref{fig:ChiralValisewFields}, there written in valise form of one row of fermions over one row of bosons.  The cis-Adinkra has SUSY enantiomer numbers $n_c = 1$ and $ n_t = 0$.
\begin{figure}[!b]
\centering
\psfrag{f1}[c][c]{
\begin{tikzpicture}
 \node[rounded rectangle,draw,fill=white!30]{$\psi_1$};
 \end{tikzpicture}~}
\psfrag{f2}[c][c]{
\begin{tikzpicture}
 \node[rounded rectangle,draw,fill=white!30]{$\psi_2$};
 \end{tikzpicture}~}
\psfrag{f3}[c][c]{
\begin{tikzpicture}
 \node[rounded rectangle,draw,fill=white!30]{$\psi_3$};
 \end{tikzpicture}~}
\psfrag{f4}[c][c]{
\begin{tikzpicture}
 \node[rounded rectangle,draw,fill=white!30]{$-\psi_4$};
 \end{tikzpicture}~}
\psfrag{b1}[c][c]{\fcolorbox{black}{white}{$A$}}
\psfrag{b4}[c][c]{\fcolorbox{black}{white}{$B$}}
\psfrag{b2}[c][c]{\fcolorbox{black}{white}{$\int dt F$}}
\psfrag{b3}[c][c]{\fcolorbox{black}{white}{$-\int dt G$}}
\includegraphics[width = 0.8 \columnwidth]{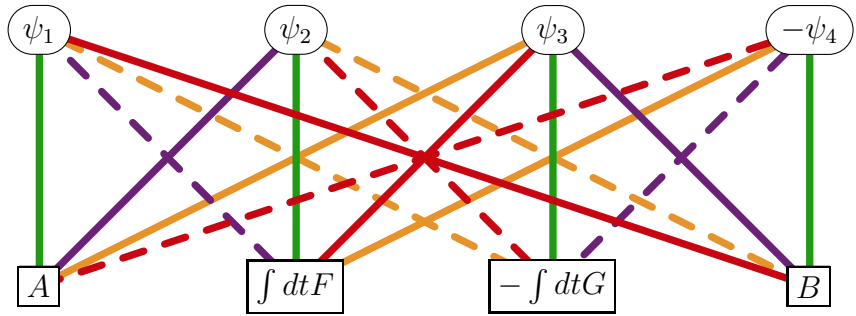}
\caption{The cis-Adinkra with nodes identified with the fields of the chiral multiplet.}
\label{fig:ChiralValisewFields}
\end{figure}
Identifying the nodes in Fig.~\ref{fig:ChiralValisewFields} as
\be
    \begin{array}{llll}
     i \Psi_1 = \psi_1, & i \Psi_2 = \psi_2, & \Psi_3 = \psi_3, & \Psi_4 = - \psi_4 \\
     \Phi_1 = A & \Phi_2 = \int~dt~F & \Phi_3 = - \int~dt~G & \Phi_4 = B
    \end{array}
\ee
we can succinctly write the zero-brane transformation laws of the chiral multiplet as:
\be
   { \rm D}_{\rm I} \Phi_i = i ( {\rm L }_{\rm I})_{i\hat{j}} \Psi_{\hat{j}},~~~{ \rm D}_{\rm I} \Psi_{\hat{j}} = ({\rm R }_{\rm I})_{\hat{j}i}\Phi_i' \nonumber\tag{\ref{eq:AdinkraMatrixDefinition}}
\ee
where the Adinkra matrices are the cis-Adinkra matrices in Eq.~(\ref{eq:cisAdinkraMatrices}), and
\be
   {\bm {\rm R}}_{\rm I} = {\bm {\rm L}}_{\rm I}^t = {\bm {\rm L}}_{\rm I}^{-1}.
\ee

\subsection{The \texorpdfstring{$4D$}{4D} \texorpdfstring{$\cal N$ =1}{N=1} Vector Multiplet}
The zero-brane reduced SUSY transformation laws for the vector multiplet, which are a symmetry of the Lagrangian~(\ref{eq:VMLagrangian0}), are:
\be
\begin{array}{llll}
 {\rm D}_1 A_{1} =  \l_2  &  {\rm D}_2A_{1}=\l_1   & {\rm D}_3A_{1}=\l_4 &   
{\rm D}_4A_{1}= \l_3~~~~~~~ \\
 {\rm D}_1 A_{2} = -  \l_4  &  {\rm D}_2A_{2}=\l_3    &{\rm D}_3A_{2}= \l_2 
  &   {\rm D}_4A_{2}= - \l_1 \\
 {\rm D}_1 A_{3} =     \l_1 &  {\rm D}_2A_{3}=-  \l_2   & {\rm D}_3A_{3}= \l_3  
 &   {\rm D}_4A_{3}=- \l_4  \\
 {\rm D}_1 {\rm d} = -  \l_3' & {\rm D}_2 {\rm d} =  -  \l_4' &  
 {\rm D}_3{\rm d}=   \l_1'  &  
{\rm D}_4{\rm d} =  \l_2' 
\end{array}
\label{eq:V1D0A}
\ee
and 
\be
\begin{array}{llll}
{\rm D}_1 \l_1 = i A_{3}' &   {\rm D}_2 \l_1 = i  A_{1}'  &  {\rm D}_3\l_1=  i  {\rm d}  
&    {\rm D}_4\l_1= - i    A_{2}' \\
{\rm D}_1 \l_2 = i  A_{1}'&   {\rm D}_2\l_2=- i  A_{3}'  &  {\rm D}_3\l_2= i   
A_{2}'&    {\rm D}_4\l_2= i {\rm d} \\
{\rm D}_1 \l_3 = - i {\rm d}  & {\rm D}_2\l_3= i   A_{2}' &  {\rm D}_3\l_3=i   
A_{3}' &    {\rm D}_4\l_3=  i  A_{1}' \\
{\rm D}_1 \l_4 = - i   A_{2}' &   {\rm D}_2\l_4=- i {\rm d} &  {\rm D}_3\l_4= i   
 A_{1}'&    {\rm D}_4\l_4= - i A_{3}'
\end{array}
\label{eq:V1D0B}
\ee
following the Adinkra construction rules as in Fig.~\ref{fig:AdinkraRules}, we arrive at the valise form for the vector multiplet Adinkra in Fig.~\ref{fig:VectorValisewFields}, which is seen to be the trans-Adinkra from~\cite{Gates:2009me}.

\begin{figure}[!h]
\centering
\psfrag{b1}[c][c]{\fcolorbox{black}{white}{$A_3$}}
  \psfrag{b2}[c][c]{\fcolorbox{black}{white}{$-A_1$}}
  \psfrag{b3}[c][c]{\fcolorbox{black}{white}{$\int {\rm d}~dt$}}
  \psfrag{b4}[c][c]{\fcolorbox{black}{white}{$-A_2$}}
  \psfrag{f1}[c][c]{
  \begin{tikzpicture}
 \node[rounded rectangle,draw,fill=white!30]{$\lambda_1$};
 \end{tikzpicture}~}
  \psfrag{f2}[c][c]{
  \begin{tikzpicture}
 \node[rounded rectangle,draw,fill=white!30]{$-\lambda_2$};
  \end{tikzpicture}~}
  \psfrag{f3}[c][c]{
  \begin{tikzpicture}
 \node[rounded rectangle,draw,fill=white!30]{$-\lambda_3$};
 \end{tikzpicture}~}
  \psfrag{f4}[c][c]{
  \begin{tikzpicture}
 \node[rounded rectangle,draw,fill=white!30]{$\lambda_4$};
 \end{tikzpicture}~}
\includegraphics[width = 0.8 \columnwidth]{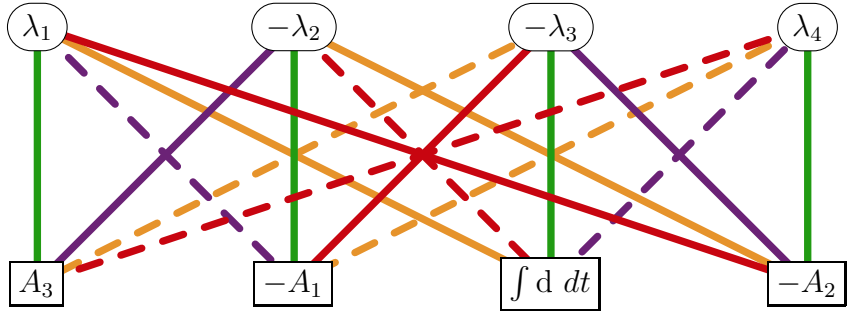}
\caption{The trans-Adinkra with nodes identified with the fields of the vector multiplet.}
\label{fig:VectorValisewFields}
\end{figure}
 Identifying the nodes in Fig.~\ref{fig:VectorValisewFields} as
\be
    \begin{array}{llll}
     i \Psi_1 = \l_1, & i \Psi_2 = -\l_2, & \Psi_3 = -\l_3, & \Psi_4 = \l_4 \\
     \Phi_1 = A_3 & \Phi_2 = -A_1 & \Phi_3 = \int{\rm d}~dt & \Phi_4 = -A_2
    \end{array}
\ee
we can succinctly write the zero-brane transformation laws of the vector multiplet as:
\be
   { \rm D}_{\rm I} \Phi_i = i ( {\rm L }_{\rm I})_{i\hat{j}} \Psi_{\hat{j}},~~~{ \rm D}_{\rm I} \Psi_{\hat{j}} = ({\rm R }_{\rm I})_{\hat{j}i}\Phi_i' \nonumber\tag{\ref{eq:AdinkraMatrixDefinition}}
\ee
where the Adinkra matrices are the trans-Adinkra matrices in Eq.~(\ref{eq:transAdinkraMatrices}), and
\be
   {\bm {\rm R}}_{\rm I} = {\bm {\rm L}}_{\rm I}^t = {\bm {\rm L}}_{\rm I}^{-1}.
\ee

\subsection{The cis- and trans-Adinkras: SUSY enantiomers}
The cis- and trans-Adinkra matrices can be succinctly written as
\be\label{eq:cisAdinkraMatrices}
       { \bm {\rm L}}_1 = {\bm {\rm I}}_4 ,~~~  { \bm {\rm L}}_2 = i {\bm \beta}^3,~~~ { \bm {\rm L}}_3 = i {\bm \beta}^2,~~~ { \bm {\rm L}}_4 = -i{\bm \beta}^1~~~\mbox{cis~~~~~}
\ee
\be\label{eq:transAdinkraMatrices}
       { \bm {\rm L}}_1 = {\bm {\rm I}}_4 ,~~~  { \bm {\rm L}}_2 = i {\bm \beta}^3,~~~ { \bm {\rm L}}_3 = -i {\bm \beta}^2,~~~ { \bm {\rm L}}_4 = -i{\bm \beta}^1~~~\mbox{trans}.
\ee
where $\beta^i$ are the $SO(4)$ generators defined in Eq.~(\ref{eq:SO4Generators}).
Here we see clearly the origination of the the names cis- and trans-: the matrices are the same, modulo a `reflection` about the `orange axis' $({\bm {\rm L}}_3)$, in an analogous fashion to how chemical \emph{enantiomers} are identical modulo a reflection about a spatial axis~\ref{fig:Chiralcarbon}.  

\section{More of the Complex Linear Superfield}
In this appendix, we explicitly show calculations pertinent to results for the complex linear multiplet reported in section~\ref{sec:CLM}.  In appendix~\ref{app:RhoZetaLinearCombinations}, we show explicitly the the SUSY transformation of a certain set of linear combinations of the complex linear multiplet.  This set of transformation laws will prove useful in appendix~\ref{app:CLMConsistencyCheck} where we show as a consistency check that, on the zero-brane, the following identity holds
\be
  {\rm D}_1 {\rm D}_2 {\rm D}_3 {\rm D_4} {\mathcal L}_{Superspace}^{(0)} =  {\mathcal L}^{(0)} + \mbox{total derivatives}
\ee
where ${\mathcal L}_{Superspace}^{(0)} = -\frac{1}{8}(K^2 + L^2)$, the Lagrangian for the complex linear multiplet in superspace, and ${\mathcal L}^{(0)}$ is the Lagrangian~\ref{eq:LagrangianCLM0} for the CLM in bosonic space-time.
\subsection{Linear Combinations of the \texorpdfstring{$4D$}{4D} \texorpdfstring{${\mathcal N}$=1}{N=1} Complex Linear Multiplet}\label{app:RhoZetaLinearCombinations}
From section~\ref{sec:CLM0}, we have following transformation laws:
$$
\begin{array}{ll}
{\rm D}_1 K= \r_1 - \z_1, & {\rm D}_2 K= \r_2 -\z_2,\\
{\rm D}_3 K=\r_3 - \z_3, & {\rm D}_4 K = \r_4 - \z_4 \\\\

 {\rm D}_1 L = -\rho _4 - \z_4 & {\rm D}_2 L = \rho _3 + \z_3\\
 {\rm D}_3 L= -\rho _2 - \z_2 & {\rm D}_4 L = \rho _1 + \z_1 \\
 \end{array}
$$
We calculate the transformation laws for the linear combinations of $\rho$ and $\zeta$ found on the right hand side:
$$
\begin{array}{ll}
{\rm {\rm D}}_1 (\rho_1 - \zeta_1) = i K' & {\rm {\rm D}}_1 (\rho_2 - \zeta_2) = -i \left(M-U_2\right) \\
{\rm {\rm D}}_1 (\rho_3 - \zeta_3) = -i \left(N-U_3\right) &{\rm {\rm D}}_1 (\rho_4 - \zeta_4) =i \left(U_0+U_1+L'\right) \\\\

{\rm {\rm D}}_2 (\rho_1 - \zeta_1)= i \left(M-U_2\right) &{\rm {\rm D}}_2 (\rho_2 - \zeta_2)= i K'\\
{\rm {\rm D}}_2 (\rho_3 - \zeta_3)= -i \left(U_0-U_1+L'\right) & {\rm {\rm D}}_2 (\rho_4 - \zeta_4)=-i \left(N+U_3\right) \\\\

{\rm {\rm D}}_3 (\rho_1 - \zeta_1)= i \left(N-U_3\right) & {\rm {\rm D}}_3 (\rho_2 - \zeta_2)= i \left(U_0-U_1+L'\right) \\
{\rm {\rm D}}_3 (\rho_3 - \zeta_3)= i K' &{\rm {\rm D}}_3 (\rho_4 - \zeta_4)= i \left(M+U_2\right) \\\\
{\rm {\rm D}}_4 (\rho_1 - \zeta_1)= -i \left(U_0+U_1+L'\right) & {\rm {\rm D}}_4 (\rho_2 - \zeta_2) =i \left(N+U_3\right) \\
{\rm {\rm D}}_4 (\rho_3 - \zeta_3)= -i \left(M+U_2\right) & {\rm {\rm D}}_4 (\rho_4 - \zeta_4)=i K'\\\\
\end{array}
$$

$$
\begin{array}{ll}
 {\rm {\rm D}}_1 (\rho_1 + \zeta_1) = i \left(V_0+V_1-K'\right) & {\rm {\rm D}}_1 (\rho_2 + \zeta_2) = -i \left(M+V_3\right) \\ {\rm {\rm D}}_1 (\rho_3 + \zeta_3) = -i
   \left(N-V_2\right) & {\rm {\rm D}}_1 (\rho_4 + \zeta_4) = -i L' \\\\
   
{\rm {\rm D}}_2 (\rho_1 + \zeta_1) =  i \left(M-V_3\right) & {\rm {\rm D}}_2 (\rho_2 + \zeta_2) = i \left(V_0-V_1-K'\right) \\ 
{\rm {\rm D}}_2 (\rho_3 + \zeta_3) =  i L' & {\rm {\rm D}}_2 (\rho_4 + \zeta_4) =  -i
   \left(N-V_2\right) \\\\

{\rm {\rm D}}_3 (\rho_1 + \zeta_1) =  i \left(N+V_2\right) & {\rm {\rm D}}_3 (\rho_2 + \zeta_2) = -i L' \\
{\rm {\rm D}}_3 (\rho_3 + \zeta_3) =  i \left(V_0-V_1-K'\right) & {\rm {\rm D}}_3 (\rho_4 + \zeta_4) = i
   \left(M+V_3\right) \\\\
   
{\rm {\rm D}}_4 (\rho_1 + \zeta_1) =  i L' & {\rm {\rm D}}_4 (\rho_2 + \zeta_2) =  i \left(N+V_2\right) \\
{\rm {\rm D}}_4 (\rho_3 + \zeta_3) =  -i \left(M-V_3\right) & {\rm {\rm D}}_4 (\rho_4 + \zeta_4) =  i
   \left(V_0+V_1-K'\right)
\end{array}
$$
The twelve linear combinations on the right hand sides of these have the transformation laws:

$$
\begin{array}{ll}
{\rm {\rm D}}_1 \left(M-U_2\right)=\z'_2-\r'_2 &
 {\rm {\rm D}}_2 \left(M-U_2\right)=\r'_1-\z'_1 \\
 {\rm {\rm D}}_3 \left(M-U_2\right)=2 \beta _3+\z'_4 &
 {\rm {\rm D}}_4 \left(M-U_2\right)=2 \beta _4-\z'_3\\
\end{array}
$$

$$
\begin{array}{ll}
 {\rm {\rm D}}_1 \left(M+U_2\right)=2 \beta _1-\z'_2 &
 {\rm {\rm D}}_2 \left(M+U_2\right)=2 \beta _2+\z'_1 \\
 {\rm {\rm D}}_3 \left(M+U_2\right)=\r'_4-\z'_4 &
 {\rm {\rm D}}_4 \left(M+U_2\right)=\z'_3-\r'_3\\
\end{array}
$$

$$
\begin{array}{ll}
 {\rm {\rm D}}_1 \left(N-U_3\right)=\z'_3-\r'_3 &
 {\rm {\rm D}}_2 \left(N-U_3\right)=-2 \beta _3-\z'_4 \\
 {\rm {\rm D}}_3 \left(N-U_3\right)=\r'_1-\z'_1 &
 {\rm {\rm D}}_4 \left(N-U_3\right)=\z'_2-2 \beta _1 \\
\end{array}
$$

$$
\begin{array}{ll}
 {\rm {\rm D}}_1 \left(U_3+N\right)=2 \beta _4-\z'_3 &
 {\rm {\rm D}}_2 \left(U_3+N\right)=\z'_4-\r'_4 \\
 {\rm {\rm D}}_3 \left(U_3+N\right)=2 \beta _2+\z'_1 &
 {\rm {\rm D}}_4 \left(U_3+N\right)=\r'_2-\z'_2
\end{array}\\
$$

$$
\begin{array}{ll}
 {\rm {\rm D}}_1 \left(M+V_3\right)=-\z'_2-\r'_2 &
 {\rm {\rm D}}_2 \left(M+V_3\right)=2 \beta _2-\z'_1 \\
 {\rm {\rm D}}_3 \left(M+V_3\right)=\z'_4+\r'_4 &
 {\rm {\rm D}}_4 \left(M+V_3\right)=2 \beta _4+\z'_3 \\
\end{array}
$$

$$
\begin{array}{ll}
 {\rm {\rm D}}_1 \left(M-V_3\right)=2 \beta _1+\z'_2 &
 {\rm {\rm D}}_2 \left(M-V_3\right)=\z'_1+\r'_1 \\
 {\rm {\rm D}}_3 \left(M-V_3\right)=2 \beta _3-\z'_4 &
 {\rm {\rm D}}_4 \left(M-V_3\right)=-\z'_3-\r'_3 \\
\end{array}
$$

$$
\begin{array}{ll}
 {\rm {\rm D}}_1 \left(V_2+N\right)=2 \beta _4+\z'_3 &
 {\rm {\rm D}}_2 \left(V_2+N\right)=\z'_4-2 \beta _3 \\
 {\rm {\rm D}}_3 \left(V_2+N\right)=\z'_1+\r'_1 &
 {\rm {\rm D}}_4 \left(V_2+N\right)=\z'_2+\r'_2\\
\end{array}
$$

$$
\begin{array}{ll}
 {\rm {\rm D}}_1 \left(N-V_2\right)=-\z'_3-\r'_3 &
 {\rm {\rm D}}_2 \left(N-V_2\right)=-\z'_4-\r'_4 \\
 {\rm {\rm D}}_3 \left(N-V_2\right)=2 \beta _2-\z'_1 &
 {\rm {\rm D}}_4 \left(N-V_2\right)=-2 \beta _1-\z'_2 \\
\end{array}
$$

$$
   \begin{array}{ll}
 {\rm {\rm D}}_1 \left(-K'+V_0+V_1\right)=\z'_1+\r'_1 &
 {\rm {\rm D}}_2 \left(-K'+V_0+V_1\right)=-2 \beta _1-\z'_2 \\
 {\rm {\rm D}}_3 \left(-K'+V_0+V_1\right)=-2 \beta _4-\z'_3 &
 {\rm {\rm D}}_4 \left(-K'+V_0+V_1\right)=\z'_4+\r'_4 \\
\end{array}
$$

$$
\begin{array}{ll}
 {\rm {\rm D}}_1 \left(-K'+V_0-V_1\right)=2 \beta _2-\z'_1 &
 {\rm {\rm D}}_2 \left(-K'+V_0-V_1\right)=\z'_2+\r'_2 \\
 {\rm {\rm D}}_3 \left(-K'+V_0-V_1\right)=\z'_3+\r'_3 &
 {\rm {\rm D}}_4 \left(-K'+V_0-V_1\right)=2 \beta _3-\z'_4 \\
\end{array}
$$

$$
\begin{array}{ll}
 {\rm {\rm D}}_1 \left(L'+U_0+U_1\right)=\r'_4-\z'_4 &
 {\rm {\rm D}}_2 \left(L'+U_0+U_1\right)=2 \beta _4-\z'_3 \\
 {\rm {\rm D}}_3 \left(L'+U_0+U_1\right)=\z'_2-2 \beta _1 &
 {\rm {\rm D}}_4 \left(L'+U_0+U_1\right)=\z'_1-\r'_1
\end{array}
$$

$$
\begin{array}{ll}
 {\rm {\rm D}}_1 \left(L'+U_0-U_1\right)=2 \beta _3+\z'_4 &
 {\rm {\rm D}}_2 \left(L'+U_0-U_1\right)=\z'_3-\r'_3 \\
 {\rm {\rm D}}_3 \left(L'+U_0-U_1\right)=\r'_2-\z'_2 &
 {\rm {\rm D}}_4 \left(L'+U_0-U_1\right)=-2 \beta _2-\z'_1\\
\end{array}
$$

This is a total of 48 equations.  Half of these uncover eight new linear combinations of fermions, these equations are:

$$
\begin{array}{ll}
 {\rm {\rm D}}_3 \left(M-U_2\right)=2 \beta _3+\z'_4 &
 {\rm {\rm D}}_4 \left(M-U_2\right)=2 \beta _4-\z'_3
\end{array}
$$
$$
\begin{array}{ll}
 {\rm {\rm D}}_1 \left(M+U_2\right)=2 \beta _1-\z'_2 &
 {\rm {\rm D}}_2 \left(M+U_2\right)=2 \beta _2+\z'_1
\end{array}
$$
$$
\begin{array}{ll}
 {\rm {\rm D}}_2 \left(N-U_3\right)=-2 \beta _3-\z'_4 &
 {\rm {\rm D}}_4 \left(N-U_3\right)=\z'_2-2 \beta _1 
\end{array}
$$
$$
\begin{array}{ll}
 {\rm {\rm D}}_1 \left(U_3+N\right)=2 \beta _4-\z'_3 &
 {\rm {\rm D}}_3 \left(U_3+N\right)=2 \beta _2+\z'_1
\end{array}\\
$$
$$
\begin{array}{ll}
 {\rm {\rm D}}_2 \left(M+V_3\right)=2 \beta _2-\z'_1 &
 {\rm {\rm D}}_4 \left(M+V_3\right)=2 \beta _4+\z'_3 
\end{array}
$$
$$
\begin{array}{ll}
 {\rm {\rm D}}_1 \left(M-V_3\right)=2 \beta _1+\z'_2 &
 {\rm {\rm D}}_3 \left(M-V_3\right)=2 \beta _3-\z'_4 
\end{array}
$$
$$
\begin{array}{ll}
 {\rm {\rm D}}_1 \left(V_2+N\right)=2 \beta _4+\z'_3 &
 {\rm {\rm D}}_2 \left(V_2+N\right)=\z'_4-2 \beta _3 
\end{array}
$$
$$
\begin{array}{ll}
 {\rm {\rm D}}_3 \left(N-V_2\right)=2 \beta _2-\z'_1 &
 {\rm {\rm D}}_4 \left(N-V_2\right)=-2 \beta _1-\z'_2 
\end{array}
$$
$$
   \begin{array}{ll}
 {\rm {\rm D}}_2 \left(-K'+V_0+V_1\right)=-2 \beta _1-\z'_2 &
 {\rm {\rm D}}_3 \left(-K'+V_0+V_1\right)=-2 \beta _4-\z'_3 
\end{array}
$$
$$
\begin{array}{ll}
 {\rm {\rm D}}_1 \left(-K'+V_0-V_1\right)=2 \beta _2-\z'_1 &
 {\rm {\rm D}}_4 \left(-K'+V_0-V_1\right)=2 \beta _3-\z'_4 
\end{array}
$$
$$
\begin{array}{ll}
 {\rm {\rm D}}_2 \left(L'+U_0+U_1\right)=2 \beta _4-\z'_3 &
 {\rm {\rm D}}_3 \left(L'+U_0+U_1\right)=\z'_2-2 \beta _1 
\end{array}
$$
$$
\begin{array}{ll}
 {\rm {\rm D}}_1 \left(L'+U_0-U_1\right)=2 \beta _3+\z'_4 &
 {\rm {\rm D}}_4 \left(L'+U_0-U_1\right)=-2 \beta _2-\z'_1  
\end{array}
$$

The ${\rm {\rm D}}_a$ transformations of these eight new fermionic fields are:

$$
\begin{array}{ll}
 {\rm D}_1\left(2 \beta _1+\z'_2\right)=i M'-i V'_3 & {\rm D}_2\left(2 \beta _1+\z'_2\right)=i K''-i V'_0-i
   V'_1 \\
 {\rm D}_3\left(2 \beta _1+\z'_2\right)=-3 i L''-2 i U'_0 & {\rm D}_4\left(2 \beta _1+\z'_2\right)=i V'_2-i N'
\end{array}
$$

$$
\begin{array}{ll}
 {\rm D}_1\left(2 \beta _1-\z'_2\right)=i M'+i U'_2 & {\rm D}_2\left(2 \beta _1-\z'_2\right)=3 i K''-2 i V'_0 \\
 {\rm D}_3\left(2 \beta _1-\z'_2\right)=-i L''-i U'_0-i U'_1 & {\rm D}_4\left(2 \beta _1-\z'_2\right)=i
   U'_3-i N'
\end{array}
$$

$$
\begin{array}{ll}
 {\rm D}_1 \left(2 \beta _2+\z'_1\right)=2 i V'_0-3 i K'' & {\rm D}_2 \left(2 \beta _2+\z'_1\right)=i M'+i U'_2
   \\
 {\rm D}_3 \left(2 \beta _2+\z'_1\right)=i U'_3+i N' & {\rm D}_4 \left(2 \beta _2+\z'_1\right)=-i L''-i U'_0+i
   U'_1
\end{array}
$$

$$
\begin{array}{ll}
 {\rm D}_1 \left(2 \beta _2-\z'_1\right)=-i K''+i V'_0-i V'_1 & {\rm D}_2 \left(2 \beta _2-\z'_1\right)=i M'+i
   V'_3 \\
 {\rm D}_3 \left(2 \beta _2-\z'_1\right)=i N'-i V'_2 & {\rm D}_4 \left(2 \beta _2-\z'_1\right)=-3 i L''-2 i U'_0
\end{array}
$$

$$
\begin{array}{ll}
 {\rm D}_1 \left(2 \beta _3+\z'_4\right)=i L''+i U'_0-i U'_1 & {\rm D}_2 \left(2 \beta _3+\z'_4\right)=i
   U'_3-i N' \\
 {\rm D}_3 \left(2 \beta _3+\z'_4\right)=i M'-i U'_2 & {\rm D}_4 \left(2 \beta _3+\z'_4\right)=2 i V'_0-3 i K''
\end{array}
$$

$$
\begin{array}{ll}
 {\rm D}_1 \left(2 \beta _3-\z'_4\right)=3 i L''+2 i U'_0 & {\rm D}_2 \left(2 \beta _3-\z'_4\right)=-i V'_2-i N'
   \\
 {\rm D}_3 \left(2 \beta _3-\z'_4\right)=i M'-i V'_3 & {\rm D}_4 \left(2 \beta _3-\z'_4\right)=-i K''+i V'_0-i
   V'_1
\end{array}
$$

$$
\begin{array}{ll}
 {\rm D}_1 \left(2 \beta _4+\z'_3\right)=i V'_2+i N' & {\rm D}_2 \left(2 \beta _4+\z'_3\right)=3 i L''+2 i U'_0
   \\
 {\rm D}_3 \left(2 \beta _4+\z'_3\right)=i K''-i V'_0-i V'_1 & {\rm D}_4 \left(2 \beta _4+\z'_3\right)=i M'+i
   V'_3
\end{array}
$$

$$
\begin{array}{ll}
 {\rm D}_1 \left(2 \beta _4-\z'_3\right)=i U'_3+i N' & {\rm D}_2 \left(2 \beta _4-\z'_3\right)=i L''+i U'_0+i
   U'_1 \\
 {\rm D}_3 \left(2 \beta _4-\z'_3\right)=3 i K''-2 i V'_0 & {\rm D}_4 \left(2 \beta _4-\z'_3\right)=i M'-i U'_2
\end{array}
$$

Only eight of these 24 equations contain on their right hand side one of the two new linear field combinations.  These eight equations are:

$$
\begin{array}{ll}
    {\rm D}_3\left(2 \beta _1+\z'_2\right)=-3 i L''-2 i U'_0 & {\rm D}_2\left(2 \beta _1-\z'_2\right)=3 i K''-2 i V'_0 \\
     {\rm D}_1 \left(2 \beta _2+\z'_1\right)=2 i V'_0-3 i K'' & 
     {\rm D}_4 \left(2 \beta _2-\z'_1\right)=-3 i L''-2 i U'_0\\
     {\rm D}_4 \left(2 \beta _3+\z'_4\right)=2 i V'_0-3 i K'' & 
      {\rm D}_1 \left(2 \beta _3-\z'_4\right)=3 i L''+2 i U'_0 \\
     {\rm D}_2 \left(2 \beta _4+\z'_3\right)=3 i L''+2 i U'_0
  &  {\rm D}_3 \left(2 \beta _4-\z'_3\right)=3 i K''-2 i V'_0
  \end{array}
$$

\subsection{From the CLM Superspace Lagrangian to the CLM Bosonic Space-time Lagrangian}\label{app:CLMConsistencyCheck}
In this section, we show that the zero-brane-reduced bosonic space-time Lagrangian
\be\eqalign{
\mathcal{L}^{(0)} = &  \frac{1}{2}K'^2 + \frac{1}{2} L'^2  - \frac{1}{2}M^2 - \frac{1}{2}N^2 + \frac{1}{4} U_\mu U^{\mu} + \frac{1}{4} V_\mu V^\mu \cr
&+ i\frac{1}{2}(\z_1 \z_1' + \z_2 \z_2' + \z_3 \z_3' + \z_4 \z_4') + i (\r_2 \b_1 - \r_1 \b_2 + \r_3 \b_4 - \r_4 \b_3)
}\nonumber\tag{\ref{eq:LagrangianCLM0}}\ee
 is recovered via proper application of four zero-brane reduced supercovariant derivatives on the full superspace Lagrangian:

\be\label{eq:CLMSUSpaceLagrangian0}\eqalign{
   {\rm D}_1{\rm D}_2{\rm D}_3{\rm D}_4 {\mathcal L}_{Superspace} = &{\rm D}_1{\rm D}_2{\rm D}_3{\rm D}_4 {\mathcal L}_{Superspace} \cr
     = &-\frac{1}{8}{\rm D}_1{\rm D}_2{\rm D}_3{\rm D}_4 \left( K^2 + L^2 \right) \cr
     = & \mathcal{L}^{(0)} + \mbox{total derivatives}.
}\ee

We calculate:
\be\label{eq:DDDDLSUSY}\eqalign{
   -4{\rm D}_1{\rm D}_2{\rm D}_3{\rm D}_4 {\mathcal L}_{Superspace} = &~({\rm D}_2{\rm D}_3K)({\rm D}_1{\rm D}_4 K) - ({\rm D}_1{\rm D}_3 K) ({\rm D}_2 {\rm D}_4 K) + \cr
   &+ ({\rm D}_1 {\rm D}_2 K) ({\rm D}_3 {\rm D}_4 K) + K {\rm D}_1{\rm D}_2{\rm D}_3{\rm D}_4K
   + \cr
   &+({\rm D}_2{\rm D}_3L)({\rm D}_1{\rm D}_4 L) - ({\rm D}_1{\rm D}_3 L) ({\rm D}_2 {\rm D}_4 L) + \cr
   &+ ({\rm D}_1 {\rm D}_2 L) ({\rm D}_3 {\rm D}_4 L) + L {\rm D}_1{\rm D}_2{\rm D}_3{\rm D}_4L
   + \cr
   &+ ({\rm D}_1{\rm D}_2{\rm D}_3 K) {\rm D}_4K + ({\rm D}_3 K)({\rm D}_1{\rm D}_2{\rm D}_4 K) + \cr
   & - ({\rm D}_2K) ({\rm D}_1{\rm D}_3{\rm D}_4 K) + ({\rm D}_1 K) ({\rm D}_2{\rm D}_3{\rm D}_4K ) + \cr
   & + ({\rm D}_1{\rm D}_2{\rm D}_3 L) {\rm D}_4L + ({\rm D}_3 L)({\rm D}_1{\rm D}_2{\rm D}_4 L) + \cr
   & - ({\rm D}_2L) ({\rm D}_1{\rm D}_3{\rm D}_4 L) + ({\rm D}_1 L) ({\rm D}_2{\rm D}_3{\rm D}_4L )
   }\ee
We break the work up by first calculating the first eight bosonic terms of~Eq.~(\ref{eq:DDDDLSUSY}).  We have for the bosonic $K$ terms
\be\eqalign{
    ({\rm D}_2{\rm D}_3K)({\rm D}_1{\rm D}_4 K)=& {\rm D_2}(\r_3 - \z_3) {\rm D}_1(\r_4 -\z_4) \cr
    =&-U_1^2 + U_0^2 + L'^2 + 2 U_0 L'
}\ee
\be\eqalign{
   ({\rm D}_1{\rm D}_3 K) ({\rm D}_2 {\rm D}_4 K) =& {\rm D}_1(\r_3 - \z_3){\rm D}_2(\r_4 - \z_4) \cr
   =&U_3^2 -N^2
}\ee

\be\eqalign{
({\rm D}_1 {\rm D}_2 K) ({\rm D}_3 {\rm D}_4 K) = &{\rm D}_1 (\r_2 - \z_2) {\rm D}_3 (\r_4 - \z_4) \cr
=& M^2 - U_2^2
}\ee
\be\eqalign{
   K {\rm D}_1{\rm D}_2{\rm D}_3{\rm D}_4K =& K {\rm D}_1{\rm 
D}_2 {\rm D}_3 (\r_4 - \z_4) \cr
     =& i K {\rm D}_1 {\rm D}_2 (M + U_2) \cr
     =& i K {\rm D}_1 (2 \b_2 + \z_1') \cr
     =& K (-2 V_0' + 3 K'') \cr
     =&-3 K'^2 + 2K' V_0 + \mbox{total derivatives}
}\ee
In the last line here, we have used integration by parts.
We have for the bosonic $L$ terms
\be\eqalign{
  ({\rm D}_2{\rm D}_3L)({\rm D}_1{\rm D}_4 L) = & {\rm D}_2(-\r_2 - \z_2){\rm D}_1(\r_1 + \z_1) \cr
  =& V_0^2 - 2 V_0 K' + K'^2 - V_1^2
}\ee
\be\eqalign{
({\rm D}_1{\rm D}_3 L) ({\rm D}_2 {\rm D}_4 L) = & - {\rm D}_1(\r_2 + \z_2) {\rm D}_2(\r_1 + \z_1) \cr
 &= -M^2 + V_3^2
}\ee
\be\eqalign{
   ({\rm D}_1 {\rm D}_2 L) ({\rm D}_3 {\rm D}_4 L) =& {\rm D}_1(\r_3 + \z_3) {\rm D}_3 (\r_1 + \z_1) \cr
   =&N^2 - V_2^2
}\ee
\be\eqalign{
   L {\rm D}_1{\rm D}_2{\rm D}_3 {\rm D}_4 L = & L {\rm D}_1 {\rm D}_2 {\rm D}_3 (\r_1 + \z_1) \cr
   = & i L {\rm D}_1 {\rm D}_2 (N + V_2) \cr
   = & i L {\rm D}_1 (-2 \b_3 + \z_4') \cr
   = &2 L U_0' + 3 LL'' \cr
   = & - 2 L' U_0 - 3L'^2 + \mbox{total derivatives}
}\ee
In the last line we have again used integration by parts.
Putting it all together we have
\be\label{eq:bosonicsummary}\eqalign{
    {\rm D}_1 {\rm D}_2 {\rm D}_3 {\rm D}_4 {\mathcal L}_{Superspace}^{bosons} = & -\frac{1}{4}\bigg(-U_1^2 + U_0^2 + L'^2 + 2 U_0 L' - U_3^2 + N^2 +  \cr
    & + M^2 - U_2^2 - 3 K'^2 + 2 K' V_0 + V_0^2 - 2 V_0 K' + \cr
    &+ K'^2 - V_1^2 + M^2 - V_3^2 +N^2 - V_2^2 - 2L' U_0 - 3L'^2 \bigg) \cr
    &+ \mbox{total derivatives} \cr
    =& \frac{1}{2} L'^2 + \frac{1}{2}K'^2 - \frac{1}{2}M^2 - \frac{1}{2}N^2 + \frac{1}{4} U_\mu U^{\mu} + \frac{1}{4} V_\mu V^\mu  + \cr
    & + \mbox{total derivatives}
}\ee
which is the bosonic part of the Lagrangian~(\ref{eq:LagrangianCLM0}).  

Next, we compute the eight fermionic parts of Eq.~(\ref{eq:DDDDLSUSY})
\be\eqalign{
   ({\rm D}_1{\rm D}_2{\rm D}_3 K) {\rm D}_4 K = &({\rm D}_1 {\rm D}_2(\r_3 - \z_3))(\r_4 - \z_4) \cr
   = & -i ({\rm D}_1(U_0 - U_1 + L'))(\r_4 - \z_4) \cr
   = & i (2 \b_3 + \z_4')(-\r_4 + \z_4) \cr
   = & i (2 \r_4 \b_3 - 2\z_4 \b_3 + \r_4 \z_4' - \z_4 \z_4')
}\ee
\be\eqalign{
({\rm D}_3 K)({\rm D}_1{\rm D}_2{\rm D}_4 K) = & (\r_3 - \z_3){\rm D}_1 {\rm D}_2 (\r_4 - \z_4) \cr
  =& -i (\r_3 - \z_3) {\rm D}_1 (N + U_3) \cr
  =& i(\r_3 - \z_3)(-2 \b_4 + \z_3') \cr
  =& i (-2 \r_3 \b_4 + 2 \z_3 \b_4 + \r_3 \z_3' - \z_3 \z_3')
}\ee

\be\eqalign{
    ({\rm D}_2 K)({\rm D}_1 {\rm D}_3 {\rm D}_4 K) =& (\r_2 - \z_2) ({\rm D}_1 {\rm D}_3 (\r_4 - \z_4)) \cr
     =& i (\r_2 - \z_2){\rm D}_1 (M + U_2) \cr
     =& i (\r_2 - \z_2) (2 \b_1 - \z_2') \cr
     =& i (2 \r_2 \b_1 - 2 \z_2 \b_1 - \r_2 \z_2' + \z_2 \z_2')
}\ee
\be\eqalign{
  ({\rm D}_1K) ({\rm D}_2 {\rm D}_3 {\rm D}_4 K) = & (\r_1 - \z_1) {\rm D}_2 {\rm D}_3 (\r_4 - \z_4) \cr
     = & i (\r_1 - \z_1 ) {\rm D}_2 (M + U_2) \cr
     = & i (\r_1 - \z_1) (2 \b_2 + \z_1') \cr
     = & i (2 \r_1 \b_2 - 2 \z_1 \b_2 + \r_1 \z_1' - \z_1 \z_1')
}\ee
\be\eqalign{
  ({\rm D}_1 {\rm D}_2 {\rm D}_3 L) ({\rm D}_4 L) = & - ({\rm D}_1 {\rm D}_2 (\r_2 + \z_2))(\r_1 + \z_1) \cr
   =& i ({\rm D}_1 (V_0 - V_1 - K'))(-\r_1 - \z_1) \cr
   =& i (2 \b_2 - \z_1')(-\r_1 - \z_1) \cr
   =& i (2 \r_1 \b_2 + 2 \z_1 \b_2 - \r_1 \z_1' - \z_1 \z_1')
}\ee
\be\eqalign{
   ({\rm D}_3L) ({\rm D}_1 {\rm D}_2 {\rm D}_4 L) = & (-\r_2 - \z_2)({\rm D}_1 {\rm D}_2(\r_1 + \z_1)) \cr
     =& i (- \r_2 - \z_2){\rm D}_1 (M - V_3) \cr
     = & i(-\r_2 - \z_2) (2 \b_1 + \z_2') \cr
     = & i (- 2\r_2 \b_1 - 2 \z_2 \b_1 - \r_2 \z_2' - \z_2 \z_2')
}\ee
\be\eqalign{
 ({\rm D}_2L)({\rm D}_1 {\rm D}_3 {\rm D}_4 L) = & (\r_3 + \z_3){\rm D}_1 {\rm D}_3 (\r_1 + \z_1) \cr
 =& i (\r_3 + \z_3) {\rm D}_1 (N + V_2) \cr
 = & i (\r_3 + \z_3) (2 \b_4 + \z_3') \cr
 = & i (2 \r_3 \b_4 + 2 \z_3 \b_4 + \r_3 \z_3'  + \z_3 \z_3')
}\ee
\be\eqalign{
  ({\rm D}_1 L) ({\rm D}_2 {\rm D}_3 {\rm D}_4 L) = & (- \r_4 - \z_4) {\rm D}_2 {\rm D}_3 (\r_1 + \z_1) \cr
  = & i (-\r_4 - \z_4) {\rm D}_2 (N + V_2 ) \cr
  =& i (\r_4 + \z_4) (2 \b_3 -\z_4') \cr
  = & i (2 \r_4 \b_3 + 2 \z_4 \b_3 - \r_4 \z_4' - \z_4 \z_4')
}\ee
All together now we have for the fermions:
\be\label{eq:fermioincalc}\eqalign{
{\rm D}_1 {\rm D}_2 {\rm D}_3 {\rm D}_4  {\mathcal L}_{Superspace}^{fermions} = & -i\frac{1}{4} \left( 2 \r_4 \b_3 - 2 \z_4 \b_3 + \r_4 \z_4' - \z_4 \z_4' + \right.\cr
 &- 2 \r_3 \b_4 + 2 \z_3 \b_4 + \r_4 \z_3' - \z_3 \z_3' + \cr
 &-2 \r_2 \b_1 + + 2 \z_2 \b_1 + \r_2 \z_2' - \z_2 \z_2' + \cr
 &+ 2 \r_1 \b_2 - 2 \z_1 \b_2 +\r_1 \z_1' - \z_1 \z_1' + \cr
 & + 2 \r_1 \b_2 + 2 \z_1 \b_2 - \r_1 \z_1' - \z_1 \z_1' + \cr
 & - 2 \r_2 \b_1 - 2 \z_2 \b_1 - \r_2 \z_2' - \z_2 \z_2' + \cr
 & -2 \r_3 \b_4 - 2 \z_3 \b_4 - \r_3 \z_3' - \z_3 \z_3' + \cr
 & \left. + 2 \r_4 \b_3 + 2 \z_4 \b_3 - \r_4 \z_4' - \z_4 \z_4' \right) \cr
 = & -i \frac{1}{2}( - \z_1 \z_1 ' - \z_2 \z_2' - \z_3 \z_3' - \z_4 \z_4') + \cr
 &-i ( \r_1 \b_2 - \r_2 \b_1 + \r_4 \b_3 - \r_3 \b_4) \cr
 = & i\frac{1}{2}(\z_1 \z_1' + \z_2 \z_2' + \z_3 \z_3' + \z_4 \z_4') + \cr
  &  + i (\r_2 \b_1 - \r_1 \b_2 + \r_3 \b_4 - \r_4 \b_3)
}\ee
which is indeed the fermionic part of the Lagrangian~(\ref{eq:LagrangianCLM0}).  Putting the bosons and fermions together, we conclude that
 \be
{\rm D}_1 {\rm D}_2 {\rm D}_3 {\rm D}_4  {\mathcal L}_{Superspace} = {\mathcal L}^{(0)} + \mbox{total derivatives}.
\ee

\section{Multiplication rules for the \texorpdfstring{${\bm {\rm a}}_I$}{aI} and \texorpdfstring{${\bm {\rm h}}_{\D}$}{hD} matrices}

The matrices denoted by ${\bm {\rm a}}_I$ and ${\bm {\rm h}}_{\D}$ defined Eqs.~\ref{eq:sl3a} and \ref{eq:sl3h} form a representation of the sl(3,R) algebra
(i. \, e. traceless linear three by three matrices with real entries).  With our definitions, the multiplication of these
is given by
$$
\begin{array}{llll}
     {\bm {\rm a}}_1 {\bm {\rm a}}_1 = -\frac{2}{3}{\bm {\rm I}}_3 + \frac{1}{2}{\bm {\rm h}}_4 + \frac{1}{6} {\bm {\rm h}}_5 & , &{\bm {\rm a}}_1 {\bm {\rm a}}_2 = \frac{1}{2}{\bm {\rm a}}_3 + \frac{1}{2}{\bm {\rm h}}_3 & , \\
     {\bm {\rm a}}_2 {\bm {\rm a}}_2 = - \frac{2}{3} {\bm {\rm I}}_3 - \frac{1}{2}{\bm {\rm h}}_4 + \frac{1}{6}{\bm {\rm h}}_5 & , & {\bm {\rm a}}_2 {\bm {\rm a}}_3 = \frac{1}{2} {\bm {\rm a}}_1 + \frac{1}{2} {\bm {\rm h}}_1 & , \\
     {\bm {\rm a}}_3 {\bm {\rm a}}_3 - \frac{2}{3} {\bm {\rm I}}_3 - \frac{1}{3} {\bm {\rm h}}_5 & , & {\bm {\rm a }}_3 {\bm {\rm a}}_1 = \frac{1}{2} {\bm {\rm a}}_2 + \frac{1}{2} {\bm {\rm h}}_2 & ,
\end{array}
\label{C1}
\eqno(C.1) $$
$$
\begin{array}{llll}
    {\bm {\rm h}}_1 {\bm {\rm h}}_1 = \frac{2}{3}{\bm {\rm I}}_3 - \frac{1}{2} {\bm {\rm h}}_4 - \frac{1}{6} {\bm {\rm h}}_5 & , & {\bm {\rm h}}_1 {\bm {\rm h}}_2 = \frac{1}{2} {\bm {\rm a}}_3 + \frac{1}{2}{\bm {\rm h}}_3 & , \\
    {\bm {\rm h}}_1 {\bm {\rm h}}_3 = -\frac{1}{2} {\bm {\rm a}}_2 + \frac{1}{2}{\bm {\rm h}}_2 & , & {\bm {\rm h}}_1 {\bm {\rm h}}_4 = - \frac{1}{2} {\bm {\rm a}}_1 - \frac{1}{2}{\bm {\rm h}}_1 & , \\
    {\bm {\rm h}}_1 {\bm {\rm h}}_5 = \frac{3}{2} {\bm {\rm a}}_1 - \frac{1}{2}{\bm {\rm h}}_1 &, &~& \\
    {\bm {\rm h}}_2 {\bm {\rm h}}_2 = \frac{2}{3} {\bm {\rm I}}_3 + \frac{1}{2} {\bm {\rm h}}_4 - \frac{1}{6}{\bm {\rm h}}_5 & , & {\bm {\rm h}}_2 {\bm {\rm h}}_3 = \frac{1}{2} {\bm {\rm a}}_1 + {\bm {\rm h}}_1 & , \\
    {\bm {\rm h}}_2 {\bm {\rm h}}_4 = - \frac{1}{2} {\bm {\rm a}}_2 + \frac{1}{2}{\bm {\rm h}}_2 & , & {\bm {\rm h}}_2 {\bm {\rm h}}_5 = - \frac{3}{2}{\bm {\rm a}}_2 - \frac{1}{2} {\bm {\rm h}}_2 & , \\
    {\bm {\rm h}}_3 {\bm {\rm h}}_3 = \frac{2}{3} {\bm {\rm I}}_3 + \frac{1}{3}{\bm {\rm h}}_5 & , & {\bm {\rm h}}_3 {\bm {\rm h}}_4 = {\bm {\rm a}}_3 & ,  \\
    {\bm {\rm h}}_3 {\bm {\rm h}}_5 = {\bm {\rm h}}_3 & , &
    {\bm {\rm h}}_4 {\bm {\rm h}}_4 = \frac{2}{3} {\bm {\rm I}}_3 + \frac{1}{3}{\bm {\rm h}}_5 & , \\
    {\bm {\rm h}}_4 {\bm {\rm h}}_5 = {\bm {\rm h}}_4 & , & 
    {\bm {\rm h}}_5 {\bm {\rm h}}_5 = 2 {\bm {\rm I}}_3 - {\bm {\rm h}}_5 & .   
\end{array}
\label{C2}
\eqno(C.2)
$$
$$
\begin{array}{llll}
   {\bm {\rm a}}_1 {\bm {\rm h}}_1 = \frac{1}{2}{\bm {\rm h}}_4 - \frac{1}{2}{\bm {\rm h}}_5 & , & {\bm {\rm a}}_1 {\bm {\rm h}}_3 = - \frac{1}{2} {\bm {\rm a}}_2 + \frac{1}{2}{\bm {\rm h}}_2 & , \\
   {\bm {\rm a}}_2 {\bm {\rm h}}_1 = -\frac{1}{2}{\bm {\rm a}}_3 + \frac{1}{2}{\bm {\rm h}}_3 & , & {\bm {\rm a}}_2 {\bm {\rm h}}_3 = - \frac{1}{2} {\bm {\rm a}}_1 - \frac{1}{2}{\bm {\rm h}}_1 & , \\
   {\bm {\rm a}}_3 {\bm {\rm h}}_1 = - \frac{1}{2}{\bm {\rm a}}_2 - \frac{1}{2} {\bm {\rm h}}_2 & , & {\bm {\rm a}}_3 {\bm {\rm h}}_3 = - {\bm {\rm h}}_4 & , \\
   {\bm {\rm a}}_1 {\bm {\rm h}}_2 = - \frac{1}{2}{\bm {\rm a}}_3 - \frac{1}{2} {\bm {\rm h}}_3 & , & {\bm {\rm a}}_1 {\bm {\rm h}}_4 = - \frac{1}{2} {\bm {\rm a}}_1 - \frac{1}{2} {\bm {\rm h}}_1 & , \\
   {\bm {\rm a}}_2 {\bm {\rm h}}_2 = \frac{1}{2}{\bm {\rm h}}_4 + \frac{1}{2} {\bm {\rm h}}_5 & , & {\bm {\rm a}}_2 {\bm {\rm h}}_4 = \frac{1}{2} {\bm {\rm a}}_2 - \frac{1}{2} {\bm {\rm h}}_2 & ,\\
   {\bm {\rm a}}_3 {\bm {\rm h}}_2 = - \frac{1}{2} {\bm {\rm a}}_1 + \frac{1}{2}{\bm {\rm h}}_1 & , & {\bm {\rm a}}_3 {\bm {\rm h}}_4 = {\bm {\rm h}}_3 & , \\
   {\bm {\rm a}}_1 {\bm {\rm h}}_5 = - \frac{1}{2} {\bm {\rm a}}_1 + \frac{3}{2} {\bm {\rm h}}_1 & , & {\bm {\rm a}}_2 {\bm {\rm h}}_5 = - \frac{1}{2} {\bm {\rm a}}_2 - \frac{3}{2}{\bm {\rm h}}_2 & , \\
   {\bm {\rm a}}_3 {\bm {\rm h}}_5 = {\bm {\rm a}}_3 & {~} & {~}&
\end{array}
\label{C.3}
\eqno(C.3)
$$

Using the facts that under the matrix transposition operator we find
$$
\left( {\bm {\rm a}}_{I} \right){}^{t}~=~ -\, {\bm {\rm a}}_{I}  ~~~,~~~  \left({\bm {\rm h}}{}_{\Delta} 
 \right){}^{t} ~=~ + \, {\bm {\rm h}}{}_{\Delta} 
\label{C.4}
\eqno(C.4)$$
the equations in (C.1) - (C.3) can easily be used to derive the commutator algebra of
all these matrices.  Some of the simplest examples of this process are illustrated below.
$$ 
\eqalign{
 &{\bm {\rm a}}_3 {\bm {\rm h}}_3 ~=~ -\, {\bm {\rm h}}_4 ~\to~  \left( {\bm {\rm a}}_3 
 {\bm {\rm h}}_3 \right){}^{t}
  ~=~ -\,  \left( {\bm {\rm h}}_4  \right){}^{t} ~\to  \cr
& 
\left( {\bm {\rm h}}_3 \right){}^{t} \left( {\bm {\rm a}}_3 \right){}^{t}
  =  - \, {\bm {\rm h}}_4  ~\to~  {\bm {\rm h}}_3  {\bm {\rm a}}_3 
  = {\bm {\rm h}}_4  ~\to  \cr
   &\left[ \, {\bm {\rm a}}_3 ~,~  {\bm {\rm h}}_3 \, \right] ~=~ -\, 2 \, {\bm {\rm h}}_4 
} \label{C.5}
\eqno(C.5)
$$
$$ 
\eqalign{
 &{\bm {\rm a}}_3 {\bm {\rm h}}_4 = {\bm {\rm h}}_3 ~\to~   \left( {\bm {\rm a}}_3 {\bm {\rm h}}_4 \right){}^{t}
  = \left( {\bm {\rm h}}_3  \right){}^{t} ~\to  \cr
& 
\left( {\bm {\rm h}}_4 \right){}^{t} \left( {\bm {\rm a}}_3 \right){}^{t}
  = {\bm {\rm h}}_3  ~\to~ - \,  {\bm {\rm h}}_4  {\bm {\rm a}}_3 
  = {\bm {\rm h}}_3  ~\to  \cr
   &\left[ \, {\bm {\rm a}}_3 ~,~  {\bm {\rm h}}_4 \, \right] ~=~ 2 \, {\bm {\rm h}}_3 
} \label{C.6}
\eqno(C.6)
$$

\newpage

\bibliographystyle{utphys}
\bibliography{Bibliography}

\end{document}